\documentclass[showkeys, reprint]{revtex4-2}
\usepackage{amsmath,amssymb,amsthm,amsfonts,graphics,graphicx,subfigure,tabularx,array}
\usepackage{makecell}
\usepackage{comment}
\usepackage{bm}
\usepackage{xcolor}

\newcommand{\ulysse}[1]{{\color{black} #1}}

\begin{document}


\title{Scale-free Points-of-Interest Distribution in a City\\ Emerging from Homogeneous Poissonian-point Processes}


\author{Eleonora Andreotti}
\email{eandreotti@fbk.eu}
\affiliation{%
 Complex Human Behaviour Lab, Fondazione Bruno Kessler, Povo (TN), Italy}%

 \author{Ulysse Marquis}%
 \affiliation{%
 Complex Human Behaviour Lab, Fondazione Bruno Kessler, Povo (TN), Italy}%
 \affiliation{Department of Mathematics, University of Trento, Via Sommarive 14, 38123 Povo (TN), Italy}
  \author{Maurizio Napolitano}%
\affiliation{%
 Digital Commons Lab, Fondazione Bruno Kessler, Povo (TN), Italy 
}%
\author{Riccardo Gallotti}%
 \affiliation{%
 Complex Human Behaviour Lab, Fondazione Bruno Kessler, Povo (TN), Italy 
}%

\date{\today}

\begin{abstract}
Urban systems often exhibit scale-invariant properties, with power-law distributions observed in various spatial and temporal patterns of human behavior. A prominent example is the distribution of commercial activities and other Points of Interest (POIs) across cities. 
However, the mechanisms by which such heavy-tailed behaviors emerge from local urban dynamics remain poorly understood. In this work, we demonstrate that global inhomogeneity in the spatial distribution of POIs can arise from the aggregation of locally homogeneous processes.

Using Foursquare data from the city of Bologna, we show that POI distributions exhibit clear power-law scaling when analyzed at city scale. 
We develop a theoretical framework in which this behavior naturally emerges from spatial clusters defined by shared intensity levels across disjoint areas, rather than spatial contiguity.
By analytically and empirically linking these local processes to the observed global distribution, we provide a generative explanation for the emergence of scale-free patterns in urban commercial structure.

To further relax the assumptions underlying the purely spatial model, and to account for the empirical observation that areas with similar activity intensity can be spatially disjoint, we introduce a hybrid hierarchical approach that combines spatial clustering with statistical heterogeneity across regions of comparable density, modeled via Poisson mixtures. This enables us to capture real-world deviations from local regularity while preserving interpretability.

Our findings highlight a key insight: complex global phenomena in cities can arise from the spatial superposition of simple, locally uniform dynamics. This connection between micro-level homogeneity and macro-scale complexity offers new tools for interpreting, modeling, and classifying urban space.
\end{abstract}

\maketitle


\section{Introduction}

\ulysse{The spatial organization of urban systems often gives rise to scale-free distributions that reflect emergent complexity and strong spatial inhomogeneity \cite{barthelemy2016structure, doi:10.1126/science.1235823, kaufmann2022scaling}. Such patterns have been extensively documented in empirical studies of urban form and mobility. A classic example is the distribution of city sizes and employment clusters, which frequently follows Zipf’s law, exhibiting power-law scaling in the upper tail (i.e., the right tail corresponding to large values) \cite{PhysRevE.87.042114, huang2018power}. Similarly, population is heterogeneously distributed across cities: a small number of large urban centers concentrate the majority of inhabitants, while many smaller cities remain sparsely populated. This rank-size pattern, associated with Zipf’s law \cite{zipf1949human}, is now known to vary across contexts and time periods, with deviations often attributable to migration dynamics or finite-time effects \cite{verbavatz2020growth}.\\
At the intra-urban scale, similar scaling behaviors emerge. The decay of population density with distance from the city center often follows an inverse-square law, resulting in a sharp concentration of activity and population near the core. This extreme centralization has been interpreted as a fractal dimension approaching zero, effectively rendering city centers as spatial singularities \cite{lemoy2017scaling}. Other models, such as those based on correlated gradient percolation theory, reproduce the fractal geometry of urban development and predict scale-free distributions in the size of built-up clusters \cite{makse1995modelling}. \\
Human mobility also displays scale-free characteristics. For example, the distribution of activity downtimes and the frequency with which individuals repeat specific mobility patterns exhibit power-law behavior \cite{bazzani2010statistical}. Alessandretti et al. \cite{alessandretti2020scales} showed that mobility trajectories are structured across nested spatial containers—such as neighborhoods, cities, and regions—with distinct characteristic sizes. Using data from over~$700{,}000$ individuals, they demonstrated that apparent scale-free behavior arises primarily through aggregation: individual mobility patterns are not themselves scale-free but become so when combined across spatial scales and populations.\\
This aggregation-based explanation is reinforced by Mizzi et al. \cite{mizzi2023}, who argue that power-law distributions in trip lengths and durations can emerge from the combination of multiple homogeneous mobility classes. Each class follows an exponential distribution consistent with the Maximum Entropy Principle, and the observed heavy tails reflect the underlying heterogeneity and multilayered structure of transportation systems rather than complex individual behavior.\\
Recent work by Gravier and Barthelemy \cite{gravier2024typologyactivitiescenturyurban} further underscores the nuanced nature of urban scaling. Analyzing nearly a century of data from Paris, they showed that different categories of urban activities, measured as the number of directory entries per category, scale differently with population. Basic services such as food retail and healthcare grow linearly, institutional services scale sublinearly, and specialized or discretionary activities, like restaurants and wine shops, scale superlinearly, reflecting innovation and consumption trends. Their analysis also captures how major historical events, such as Haussmann’s renovation or the Paris Commune, induced temporary shifts in these scaling dynamics.\\
A related perspective is offered by Malmgren et al. \cite{malmgren2009poisson, malmgren2008letter}, who explain heavy-tailed inter-event time distributions in human communication as arising from simple mechanisms: stationary Poisson processes modulated by long-term, nonstationary patterns of activity. These results echo broader themes in urban studies, suggesting that scale-free behavior can often emerge from the superposition of regular, localized processes modulated by long-range variability or aggregation.}

In this work, we analyze the distribution of POIs across six distinct categories (\texttt{Business and Professional Services}, \texttt{Community and Government}, \texttt{Dining and Drinking}, \texttt{Health and Medicine}, \texttt{Retail}, and \texttt{Travel and Transportation}). To study the spatial distribution of POIs, we define the variable number of POIs within fixed-radius regions of 50 meters. A radius of 50 meters represents a trade-off between spatial resolution and coverage: it is small enough to capture local variations in POI density, yet sufficiently large to avoid missing medium-to-high density clusters and broader spatial patterns \cite{Li_2018}. At a global scale, i.e., by aggregating data across all regions in the city without distinguishing between specific neighborhoods, the distribution of POI density exhibits power-law behavior, with exponents ranging from 2.7 to 4.5. \\
These distributions are characterized by both soft and hard cutoffs, depending on the category of the POIs. The scale-free nature of the data is linked to significant heterogeneity in the city's properties: some areas are highly developed and commercially dense, while others remain sparsely populated or underdeveloped. The varying exponents and cutoffs across categories suggest not only differences in development between areas but also distinct characteristics of the economic activities present in these locations. The only two categories where POIs follow a power-law distribution without any cutoff are \texttt{Travel and Transportation} and \texttt{Retail}. With relatively low exponents ($\alpha$ = 2.7 and $\alpha$ = 2.9, respectively), these categories suggest that it is less influenced by physical constraints and network size compared to other categories, exhibiting a more even distribution across the city \cite{gallotti2021unraveling}.\\ 
Building on these observations, a key contribution of this study is the novel connection we establish between the power-law distributions observed at the global scale and the underlying homogeneous Poisson Point Processes identified at the local scale. This perspective aligns with the view that cities are complex systems in which large-scale order arises from decentralized, bottom-up processes, as discussed by Batty in his work on the scaling and morphology of urban form \cite{Batty2008}.\\
To this end, we investigate two alternative mechanisms that can give rise to the observed power-law distributions. The first approach is based on a hierarchical clustering strategy. 
Specifically, by using hierarchical clustering techniques, we demonstrate that power-law distributions arise from spatial clusters of homogeneous Poissonian processes distributed across the city's regions. \\
Moreover, we mathematically show that a power-law distribution can emerge globally as a consequence of local Poissonian distributions, provided that the areas and intensities of these processes follow appropriate distributions. \\
This approach provides valuable insights into the processes that govern the spatial arrangement of commercial POIs, demonstrating how the power-law exponents, derived from clustering, further characterize the neighborhoods.\\
However, as we will demonstrate mathematically, this explanation relies on rather strong assumptions—such as the existence of well-defined spatial clusters and specific scaling relationships—that are unlikely to hold in complex urban environments. Empirically, the clusters contributing to the observed power-law behavior are not necessarily unique or clearly delineated; portions of one cluster may overlap with or contribute to others, effectively generating local Poisson-like distributions with varying parameters.\\
To address the limitations, we introduce a second approach based on a Poisson mixture model \cite{mclachlan200001}. Specifically, we extend our framework by incorporating a hierarchical structure in which each spatial cluster contains a Poisson mixture. This hybrid formulation retains the spatial organization observed in empirical data, while relaxing the strict assumptions required for global power-law behavior. In particular, the emergence of heavy tails no longer depends on a specific scaling of cluster areas or on a one-to-one association between clusters and intensities.\\
Instead, the global distribution arises from the aggregate contribution of multiple clusters, each potentially containing a range of intensities with heterogeneous mixture weights. By decoupling spatial size from statistical intensity, this model preserves the interpretability of spatial clustering while introducing greater flexibility in the underlying generative mechanism. Ultimately, it bridges the spatial and statistical viewpoints, offering a more general explanation for the heavy-tailed patterns observed in urban POI distributions. \\
To complement our analytical results, we introduce a generative model for creating synthetic cities with controlled structural features. We first employ this tool to validate our theoretical outcomes under the original assumptions, then progressively relax those assumptions, such as strict cluster definitions or fixed area, intensity scalings, to test the robustness of scale-free emergence under more flexible spatial and statistical configurations.\\
Building on this approach, the empirical study was carried out using data from the city of Bologna. Bologna was selected as a representative example of a medium-sized European city, combining a well-preserved historical center with diverse commercial activities distributed across distinct administrative districts. These characteristics make it a suitable testbed for validating our theoretical framework in a real-world urban environment. While Bologna is the empirical focus, both the methodology and the analytically derived mathematical results are designed to be generalizable and adaptable to other urban contexts.

\textbf{Structure of the Paper.}\\
The structure of this paper is organized to guide the reader from empirical observation to theoretical interpretation and back to validation. \\
Section~\ref{sec:met} introduces the dataset and outlines the methodological pipeline used to characterize the spatial distribution of POIs in the city of Bologna. This includes both the detection of global power-law behavior and the identification of locally homogeneous regions through hierarchical clustering based on POI density.
Section~\ref{sec:math} develops the theoretical framework linking local Poissonian processes to global heavy-tailed distributions. Two complementary models are introduced: a spatially explicit formulation based on local intensity variations, and a latent mixture model that captures heterogeneity through probabilistic assignment rather than spatial partitioning. The two perspectives are then unified through a hybrid hierarchical model that incorporates both spatial organization and latent variability.
Section~\ref{sec:val} is dedicated to the empirical validation of the theoretical results. We first generate synthetic surfaces to reproduce the theoretical assumptions under controlled conditions and observe the emergence of power-law distributions from structured Poisson processes. We then assess whether similar conditions can be found in the real POI data. By re-clustering the urban surface into artificial regions with controlled area scaling, we verify that the theoretical predictions hold. Finally, we demonstrate how the hybrid model, combining latent probabilities and Poisson intensities, offers a statistically coherent interpretation of the empirical patterns observed.\\
The paper concludes in Section~\ref{sec:con}, where we summarize the main findings, highlight their conceptual and methodological contributions, and discuss potential applications and directions for future research.

\section{Methodology}\label{sec:met} 
\subsection{Data Source}
\begin{figure}[h!]
    \centering
    \subfigure[The size of the bar describes the number of POI associated to each category in Bologna]{\includegraphics[width=.48\linewidth]{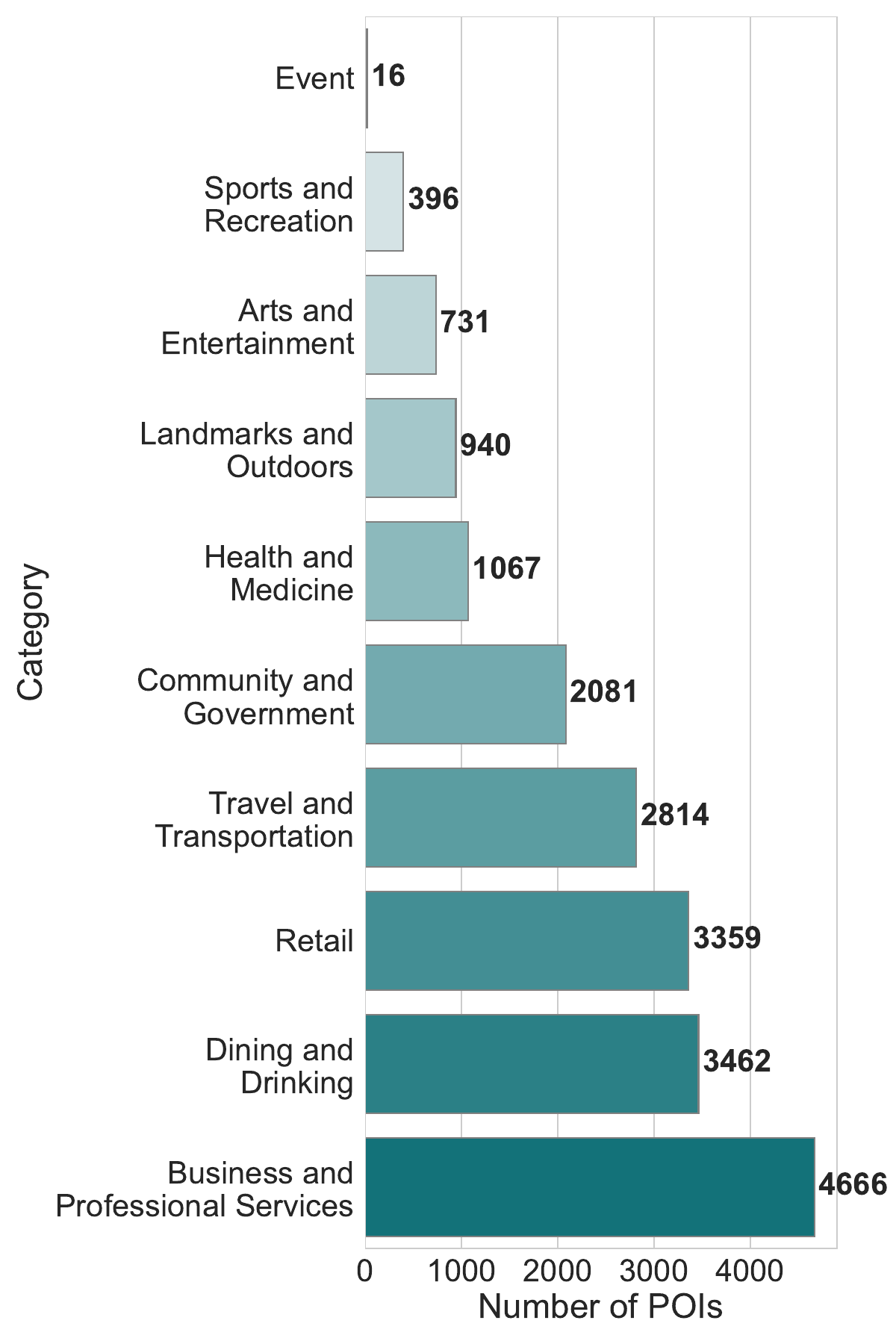}}
    \subfigure[Circle size indicates the number of POIs in a given category (row) within a district (column)]{\includegraphics[width=.50\linewidth]{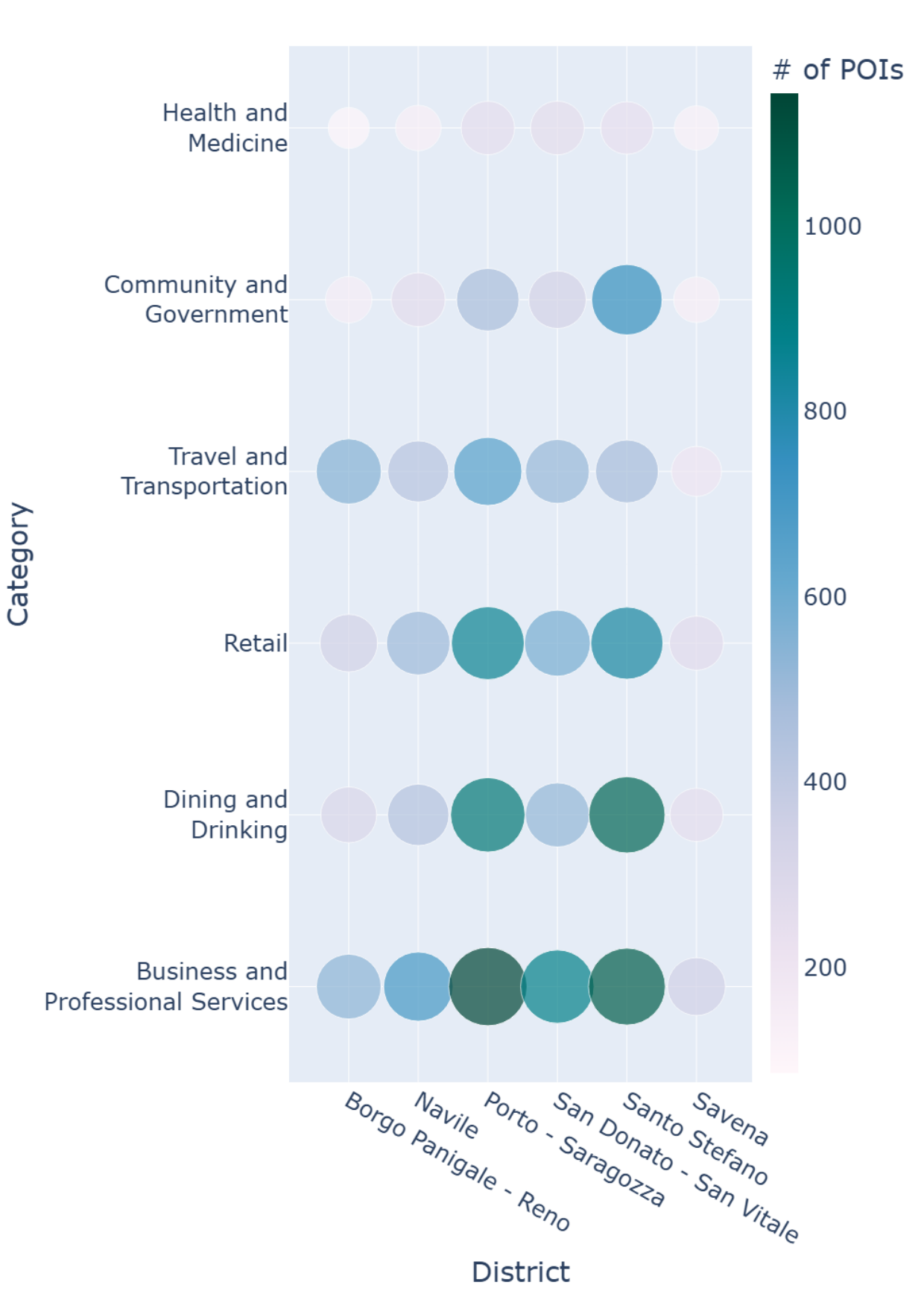}}
 \caption{Number of POIs per Foursquare category and district.}   
 \label{fig:numPOIs}
 \end{figure}

\begin{figure}[h!]
\centering
    \includegraphics[width=.8\linewidth]{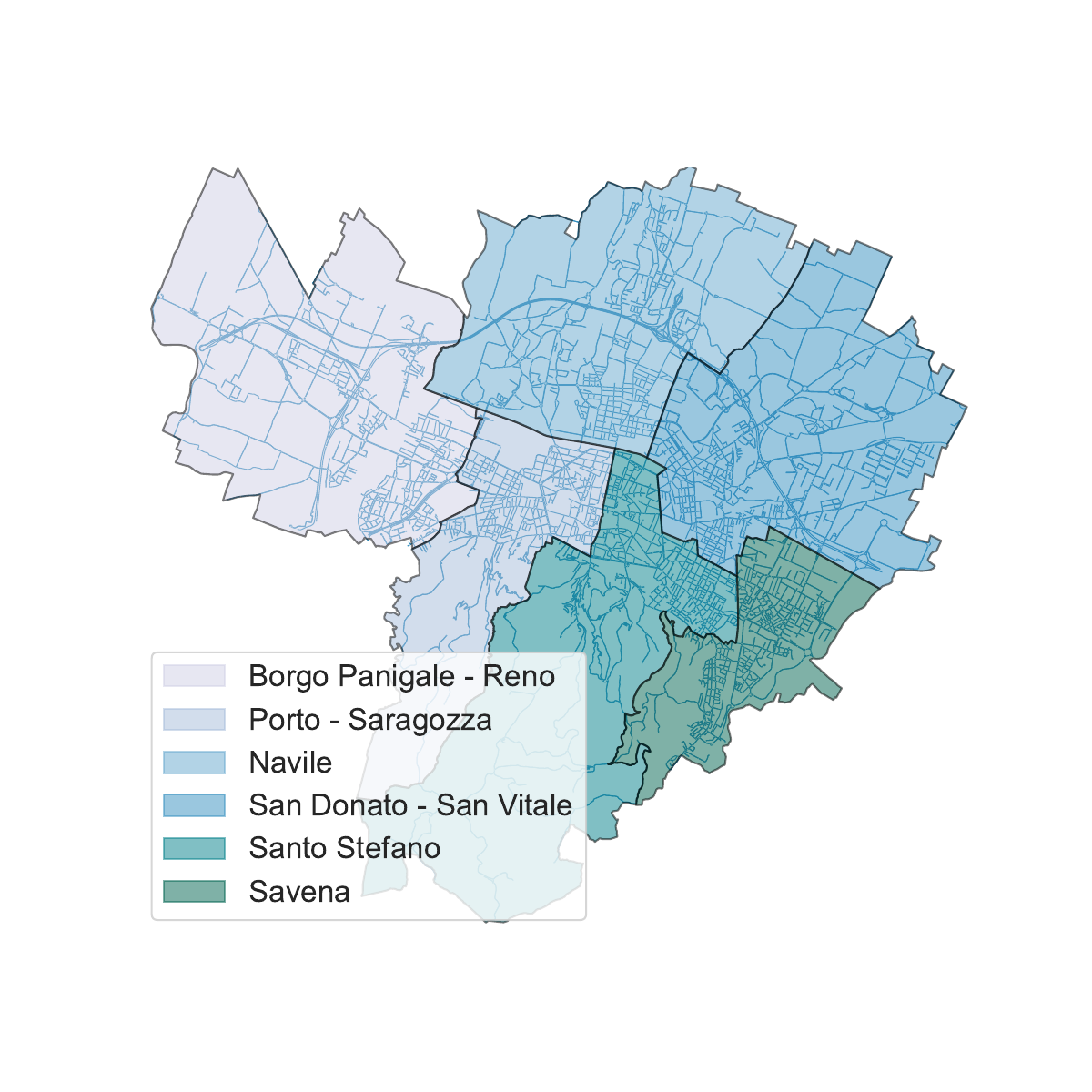}
 \caption{Visualization of Bologna’s districts, displayed with the underlying OpenStreetMap street network \cite{openstreetmap}; district areas are color-coded according to the legend. District boundary polygons are obtained from the Open Data portal of the Municipality of Bologna \cite{opendata_bologna}.}   
 \label{fig:circo}
 \end{figure}

In this work, we focus on the Points of Interest (POIs) from the Foursquare dataset~\cite{foursquare}, characterizing them based on the first level of the hierarchical category structure assigned by Foursquare, as summarized for the city of Bologna in Fig.~\ref{fig:numPOIs}(a).\\
We focus our analysis on the most prevalent POI categories in the city of Bologna, defined as those comprising at least 1,000 POIs. The Bubble Chart in Fig.~\ref{fig:numPOIs}(b) displays the number of POIs, broken down by category and district. \\
The city of Bologna, like many other medium-to-large cities, is administratively divided into districts (circoscrizioni), Fig.~\ref{fig:circo}. Each is governed by a Neighborhood Council, which represents citizens directly and plays a political, advisory, and decision-making role in local governance. These councils manage basic services and local initiatives, potentially leading to distinct characteristics and dynamics across neighborhoods ~\cite{comune_bologna_amministrazione}.\\
The districts with the highest concentration of POIs are \texttt{Santo Stefano} and \texttt{Porto-Saragozza}, though this is not consistent across all categories. For example, the \texttt{Health and Medicine} category has more POIs in the \texttt{San Donato-San Vitale} district, while the \texttt{Community and Government} category is more prevalent in the \texttt{Santo Stefano} district.\\
To better understand these patterns, a more in-depth analysis of POI distributions and district characteristics is required.

\subsection{Fitting Power Laws and Estimating the Scaling Parameters}

To identify the appropriate distribution that best fits the data, we analyze the empirical distribution of the number of POIs per spatial unit, where each unit corresponds to a geodesic ball of fixed radius (see, e.g., \cite{jost2017riemannian} for a formal definition of geodesic balls). For brevity, in what follows we shall simply refer to them as \emph{balls}.
The urban surface is tessellated into a grid of overlapping balls of radius 50 m, each serving as a sampling window for POI counts. These balls are laid out to systematically cover the entire city area, allowing us to obtain a spatially exhaustive and resolution-consistent set of count observations. The resulting dataset consists of integer-valued POI counts associated with each ball, which we then aggregate into a frequency distribution.\\
Given the pronounced heterogeneity typically observed in urban phenomena, where a few locations concentrate a large number of activities while most remain sparsely populated, it is reasonable to investigate whether the empirical distribution of POI counts exhibits power-law behavior. Such a pattern would imply the presence of scale-free organization and heavy-tailed distributions, features that are common in many spatial and socio-economic systems.\\

A power-law distribution is defined by the functional form
\begin{equation}
    P(x) \sim x^{-\alpha}
\end{equation}
where $\alpha > 1$ is the scaling exponent, and the relationship holds for $x \geq x_{\min}$. This implies that the probability of observing a value $x$ decays polynomially with $x$, in contrast to the exponential decay seen in light-tailed distributions. The idea of power-law scaling originates with Pareto's study of wealth distribution~\cite{Pareto1897}, and was later generalized by Zipf in the context of word frequencies~\cite{zipf1949human}. 
More broadly, power-law behavior is often interpreted as a hallmark of self-organized criticality: slowly driven, dissipative systems spontaneously organize to a critical state displaying scale-free avalanche dynamics~\cite{bak1996how}. 
In recent decades, power laws have also been extensively studied in complex systems and network theory, most notably as signatures of preferential attachment in scale-free networks~\cite{barabasi1999emergence}.
Identifying such a pattern in POI data can therefore shed light on the spatial mechanisms shaping urban structure and activity concentration.\\
To determine whether this distribution exhibits power-law behavior, we employ the method proposed by Clauset and collaborators \cite{Clauset}, which is widely used for testing power-law fits in empirical data, particularly when dealing with heavy-tailed or non-standard patterns \cite{malmgren2009universality, gallotti2016stochastic}.  \\
This method involves fitting a variety of candidate distributions and selecting the one that best describes the observed data based on a variety of statistical criteria. For parameter estimation, the Maximum Likelihood Estimator (MLE) is used, as it provides more accurate and robust estimates compared to other methods \cite{Goldstein04}. MLE also allows for the application of a Kolmogorov-Smirnov (KS) test, which is crucial for assessing the goodness-of-fit of the chosen distribution. The KS test compares the empirical distribution function with the distribution function of the fitted model, offering a quantitative measure of the model's adequacy in capturing the characteristics of the observed data.
If the resulting p-value is greater than~$0.1$, the power-law is a plausible hypothesis for the data, otherwise it is rejected.
For the entire dataset and for each category, we compared several common discrete distributions, including Power Law, Exponential, Truncated Power Law, Poisson, and Lognormal, to determine which best describes the observed data. The selection was based on both quantitative metrics (such as log-likelihood values and p-values from distribution comparison tests) and visual inspection of fit, such as graphical representations of the cumulative distribution function (CDF), the complementary cumulative distribution function (CCDF) and probability mass function (PMF). Given that the variable under study (number of POIs per spatial unit) is discrete, the probability mass function (PMF) is used rather than the probability density function (PDF).

\begin{figure}[h!]
    \centering
    \subfigure[]{\includegraphics[width=1\linewidth]{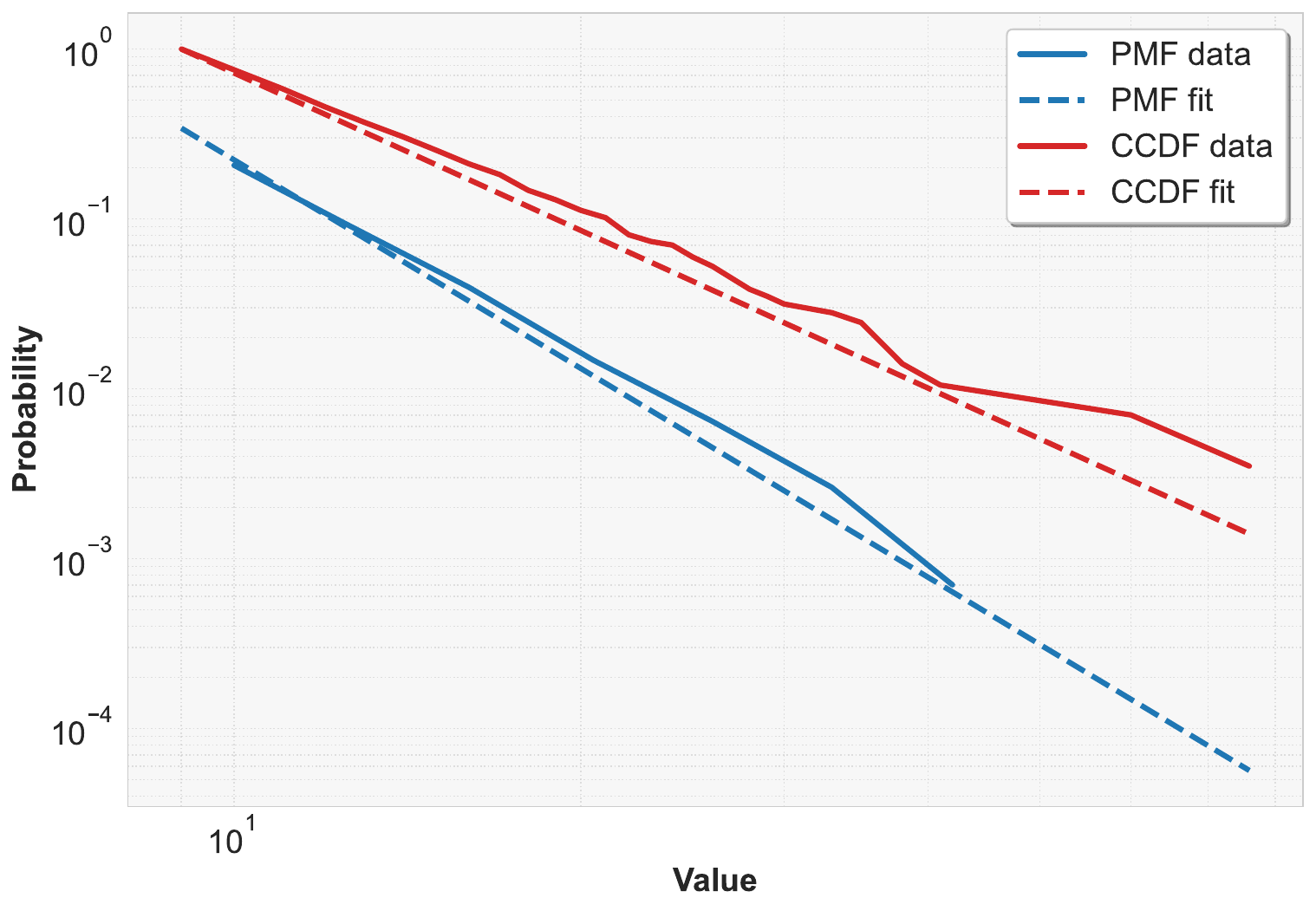}}
    \subfigure[]{\includegraphics[width=1\linewidth]{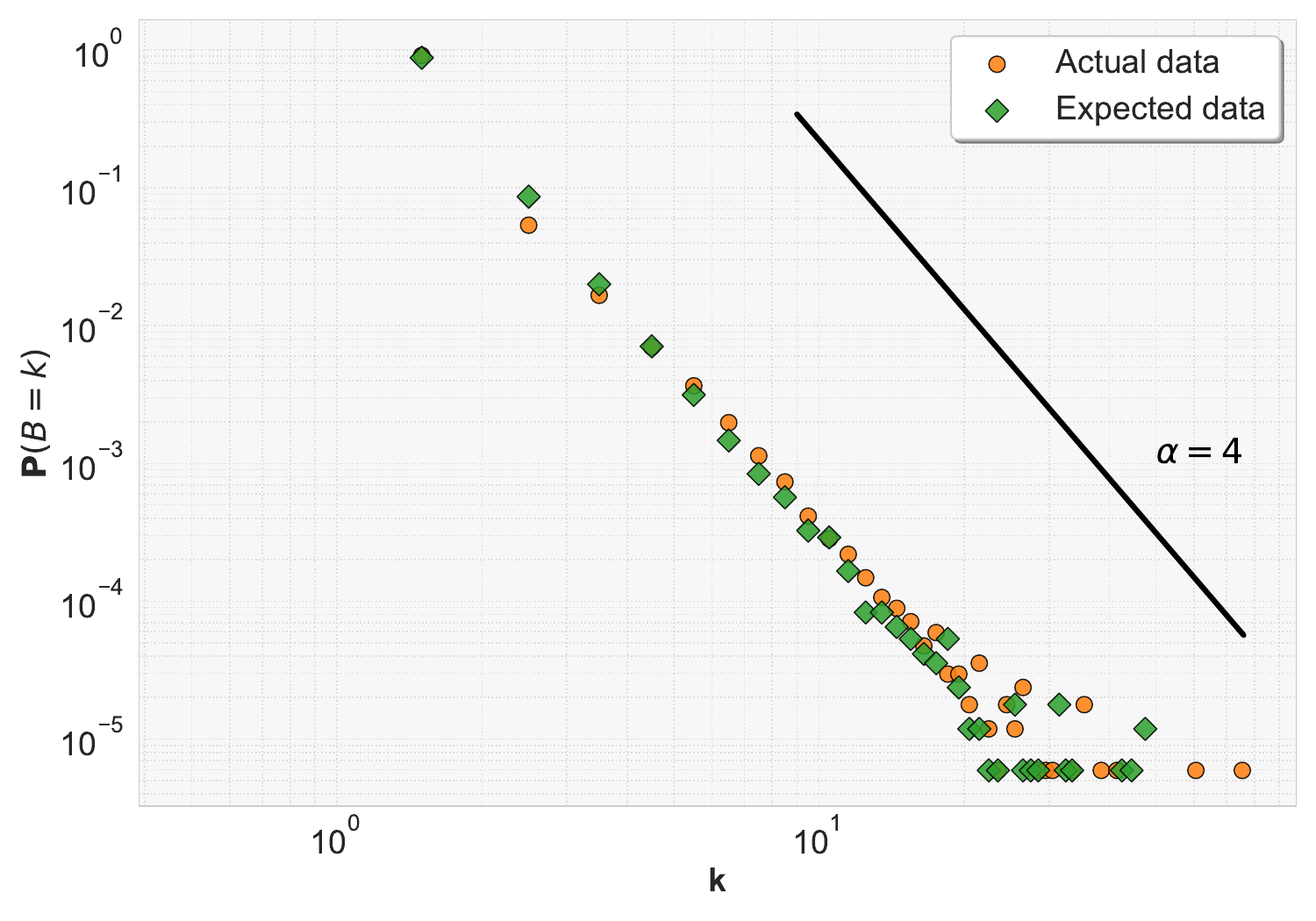}}
    \caption{Entire dataset plotted on a logarithmic scale. (a) Probability mass function (PMF) and complementary cumulative distribution function (CCDF) of the empirical data compared with the theoretical fit. The PMF is shown in red, the CCDF in blue, while dashed lines indicate the corresponding theoretical fits. (b) Histograms of POIs: empirical data are shown in orange, expected data in green. The solid black line represents a power-law with exponent $\alpha = 4$, proposed as a visual guide.}\label{fig:pl_tutte}
\end{figure}
The results on the entire dataset suggests that the actual data follows a power law distribution, where the exponent of the power law is~$\approx 4.1$, indicating how quickly the probability decreases as the observed values increase, Fig.~\ref{fig:pl_tutte}. The scale-free nature of the data distribution is associated to a large heterogeneity in the city's properties, as some areas are highly developed and densely packed, while others may be sparsely populated or less developed.
The p-value from the KS test which compares the empirical distribution with the theoretical distribution, is~$0.2929$. The high p-value indicates that there is not enough evidence to reject the null hypothesis, which in this case is that the actual data follows the theoretical distribution (in this case, the power law). 
By the likelihood ratio test (LRT), we compare the power-law model to alternative distributions, concluding that since the p-value of the KS test is greater than~$0.1$, and the power-law model is favored over the alternatives by the LRT, there is relatively strong support for the data following a power-law distribution.\\
Once the distribution of the entire dataset was analyzed, we narrowed the focus to the data by category, in order to verify whether the global power-law behavior of the data is also reflected in the individual categories.\\
We observe that the only categories that can be represented by a pure power law are the \texttt{Travel and Transportation} and \texttt{Retail} categories, with slopes of $\approx 2.7$ (K-S test:~$D = 0.0902$, p-value~$= 0.1538$) and~$\approx 2.9$ (K-S test:~$D = 0.0664$, p-value~$= 0.1191$), respectively. The other four categories, on the other hand, are characterized by both soft and hard cutoffs, depending on the category.
The scale-free nature of the data distribution is associated with significant heterogeneity in the city's properties, as some areas are highly developed and commercially dense, while others remain sparsely populated or underdeveloped. The varying exponents and cutoffs across categories suggest not only differences in development between areas but also distinct characteristics of the economic activities present in different locations.

\subsection{Empirical Observation of Local Poisson Behavior via Hierarchical Clustering}

To further explore the spatial structure underlying the distribution of POIs in the city, we apply a hierarchical variant of the DBSCAN clustering algorithm~\cite{ran2023hierarchical}. Our aim is to investigate whether spatial clusters of POIs, identified at multiple density levels, exhibit local regularities compatible with a Poisson point process.

The DBSCAN algorithm was selected as a clustering method due to its suitability for identifying spatially coherent regions of varying density without requiring prior knowledge of the number of clusters \cite{ester1996dbscan, schubert2017dbscan}. This is particularly appropriate in urban settings, where POI distributions tend to exhibit irregular and fragmented spatial patterns. Unlike methods that impose geometric or parametric constraints (for instance ~$k$-means or Gaussian mixture models, see \cite{EZUGWU2022104743}), DBSCAN is density-based and can detect arbitrarily shaped clusters while naturally excluding outliers.

In the context of our analysis, DBSCAN offers an operational definition of a ``region" as a maximal set of neighboring balls that exceed a local density threshold, specified via the parameters \texttt{epsilon}, the neighborhood radius, and \texttt{minPts}, the minimum number of points. This makes it especially well suited to uncover spatial domains that may plausibly correspond to areas governed by homogeneous Poisson processes. By iteratively applying DBSCAN with decreasing density thresholds, we isolate clusters that are locally dense and internally homogeneous, allowing us to test the hypothesis that POI counts within such regions follow a Poisson distribution with constant rate.\\
We cover the urban surface by combining a random ball covering, implemented as a Boolean model of fixed-radius balls~\citep{stoyan1995stochastic}, with a greedy ball covering step applied to the uncovered gaps. This procedure ensures a complete coverage of the study area, where each ball represents an elementary observational unit of $\approx 7{,}854 \,\text{m}^2$ (see Figure~\ref{fig:covering} in Appendix).\\
To identify locally dense regions, we apply the DBSCAN algorithm using a range of combinations of the two core parameters: \texttt{epsilon} and \texttt{minPts}. Specifically, the radius \texttt{epsilon} is varied from~$50$ meters up to~$7$ kilometers, while \texttt{minPts} is varied from a maximum of~$50$ down to a minimum of~$3$. This allows us to explore the spatial structure of POI density across multiple scales, from highly localized clusters to more spatially extended regions.

\begin{figure}[h!]
    \centering
    \subfigure[{\shortstack{Business and\\Professional Services}}]{
      \includegraphics[width=.54\linewidth]{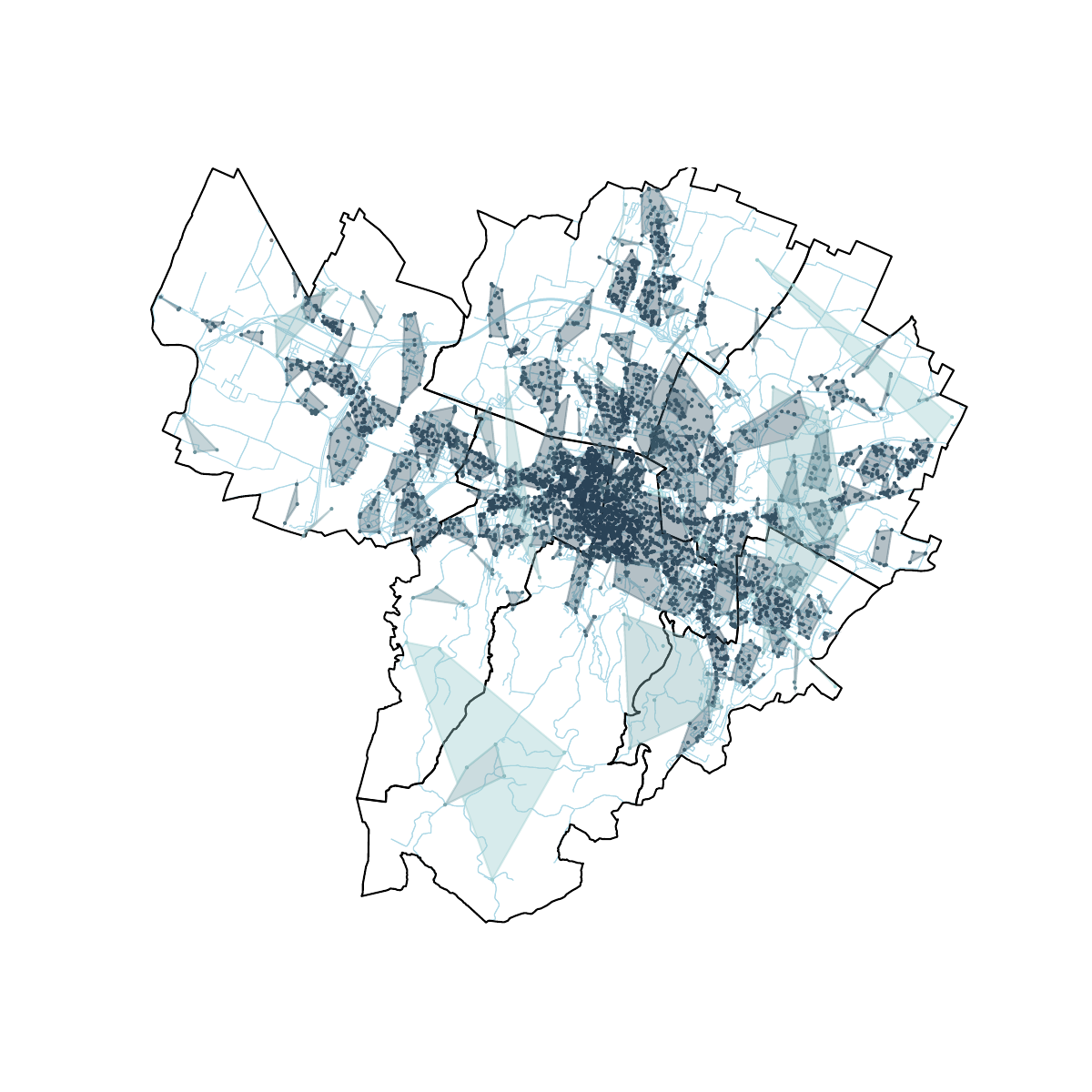}
    }\hspace{-1.3cm}
    \subfigure[{\shortstack{Community and\\Government}}]{
      \includegraphics[width=.54\linewidth]{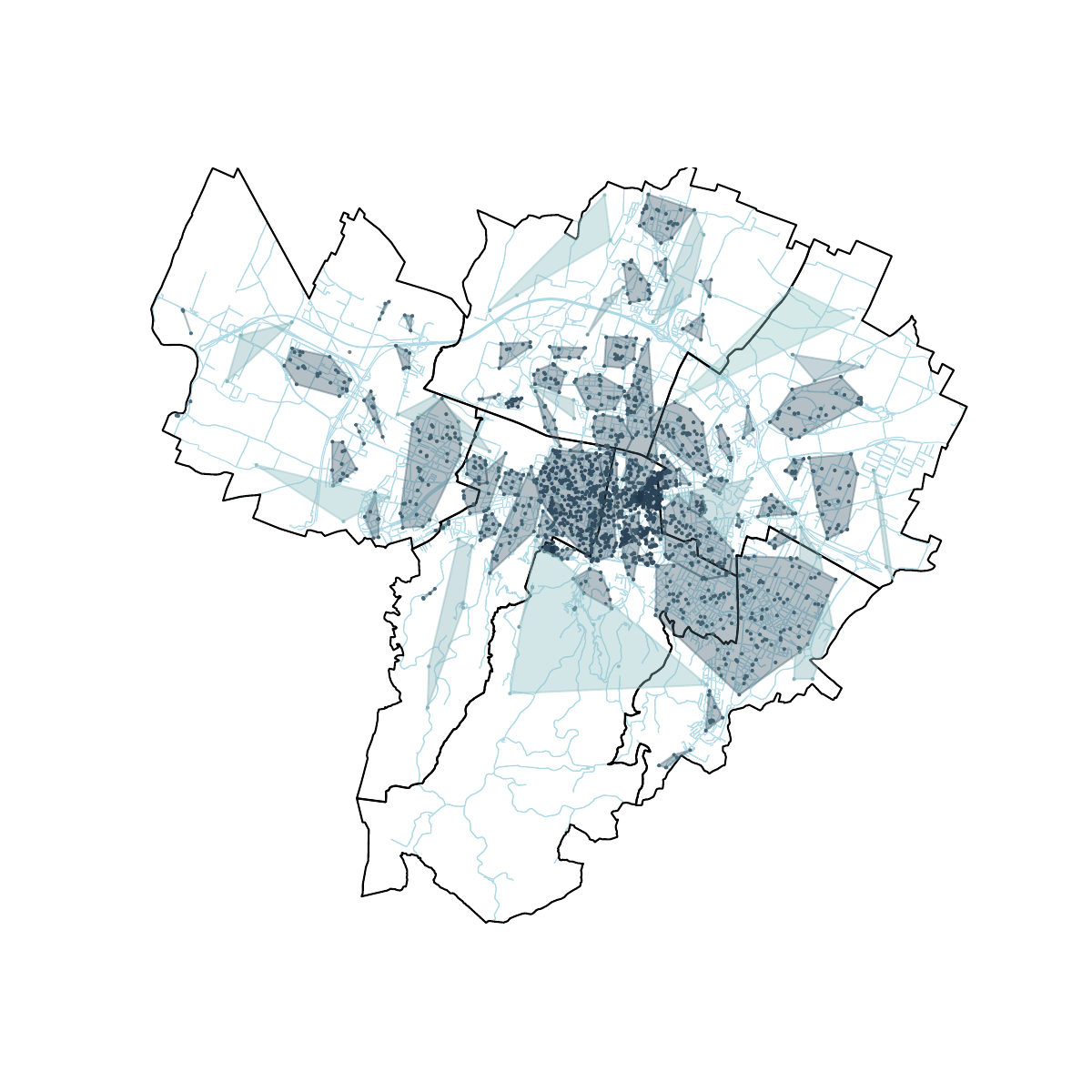}
    }
    \subfigure[Dining and Drinking]{
      \includegraphics[width=.54\linewidth]{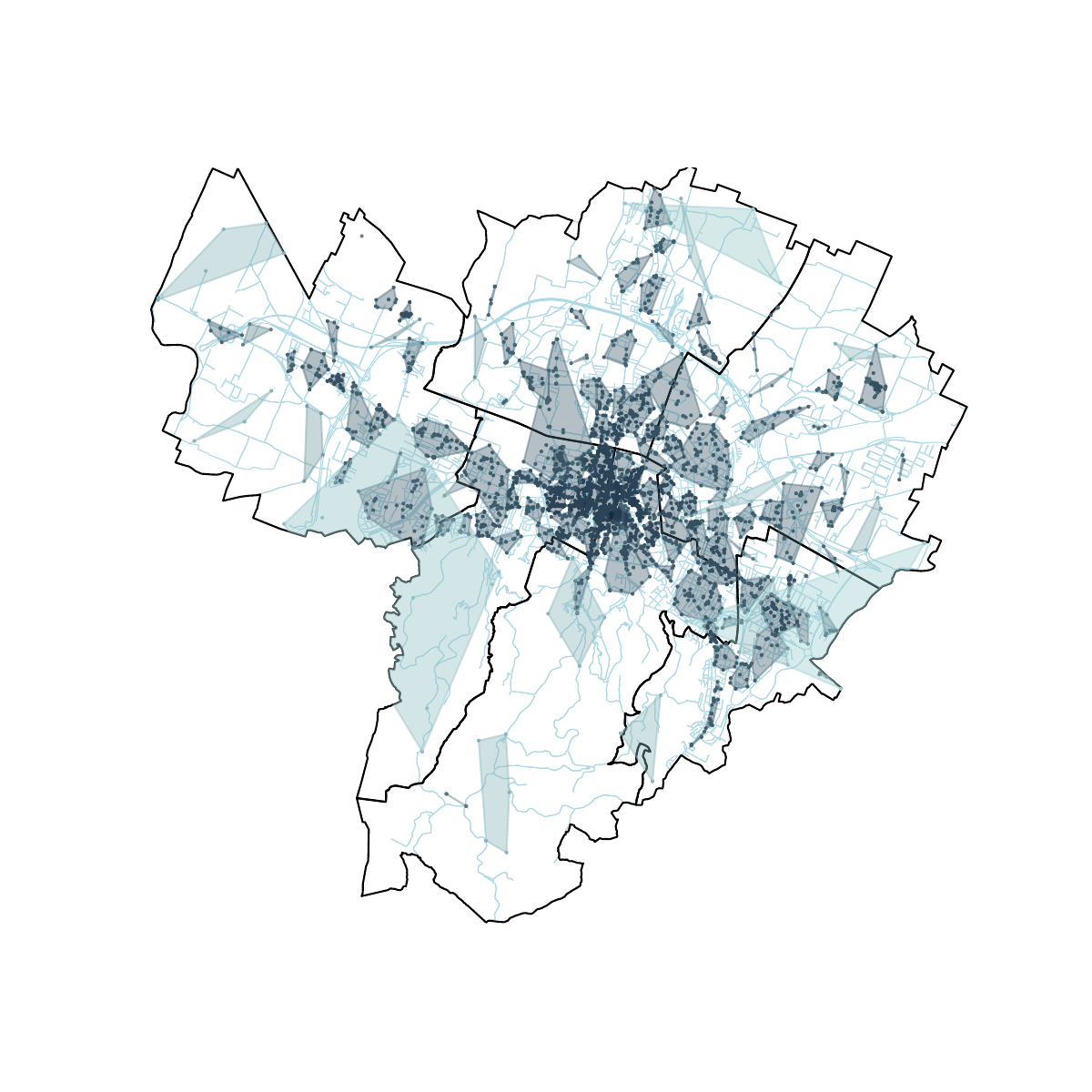}
    }\hspace{-1.3cm}
    \subfigure[Health and Medicine]{
      \includegraphics[width=.54\linewidth]{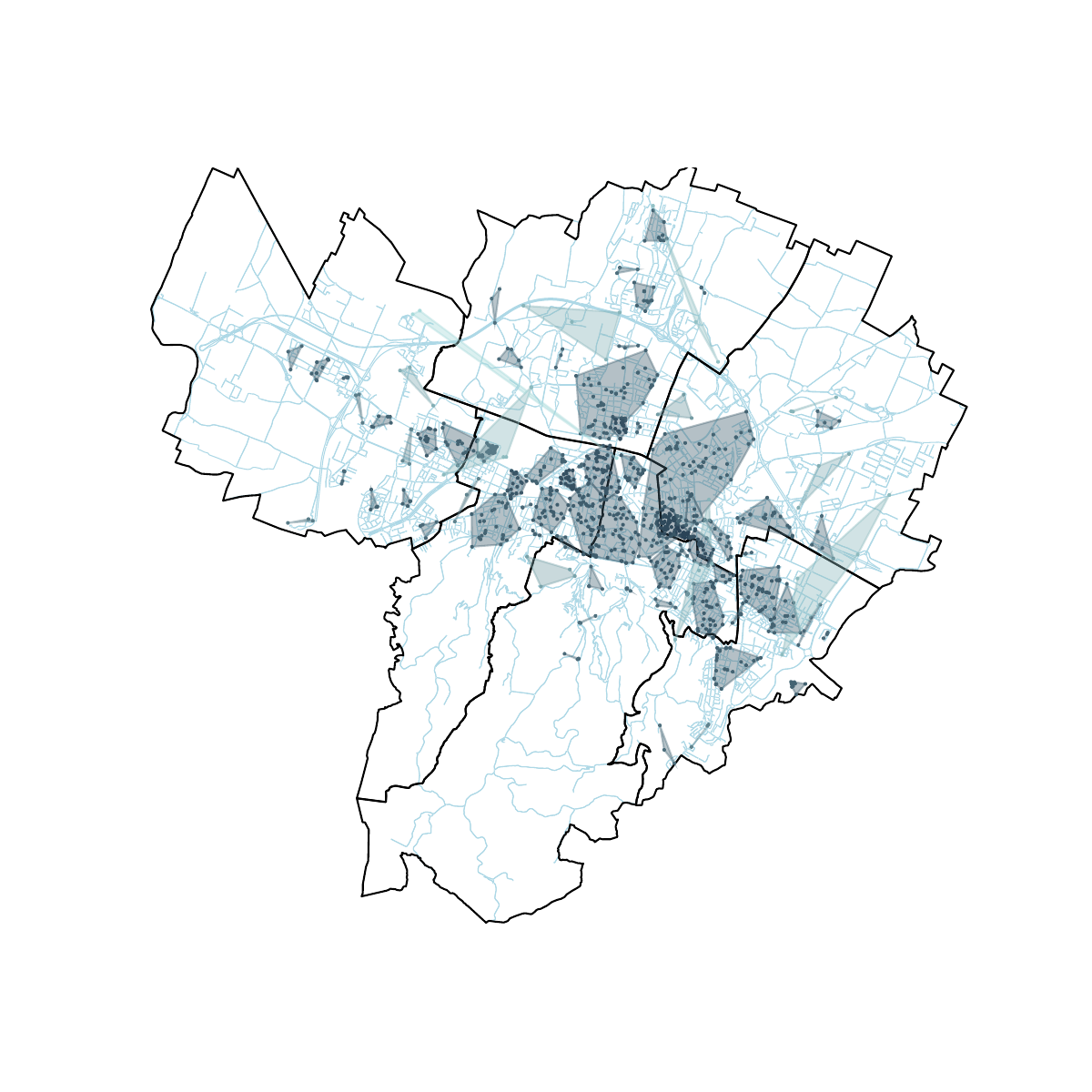}
    }
    \subfigure[Retail]{
      \includegraphics[width=.54\linewidth]{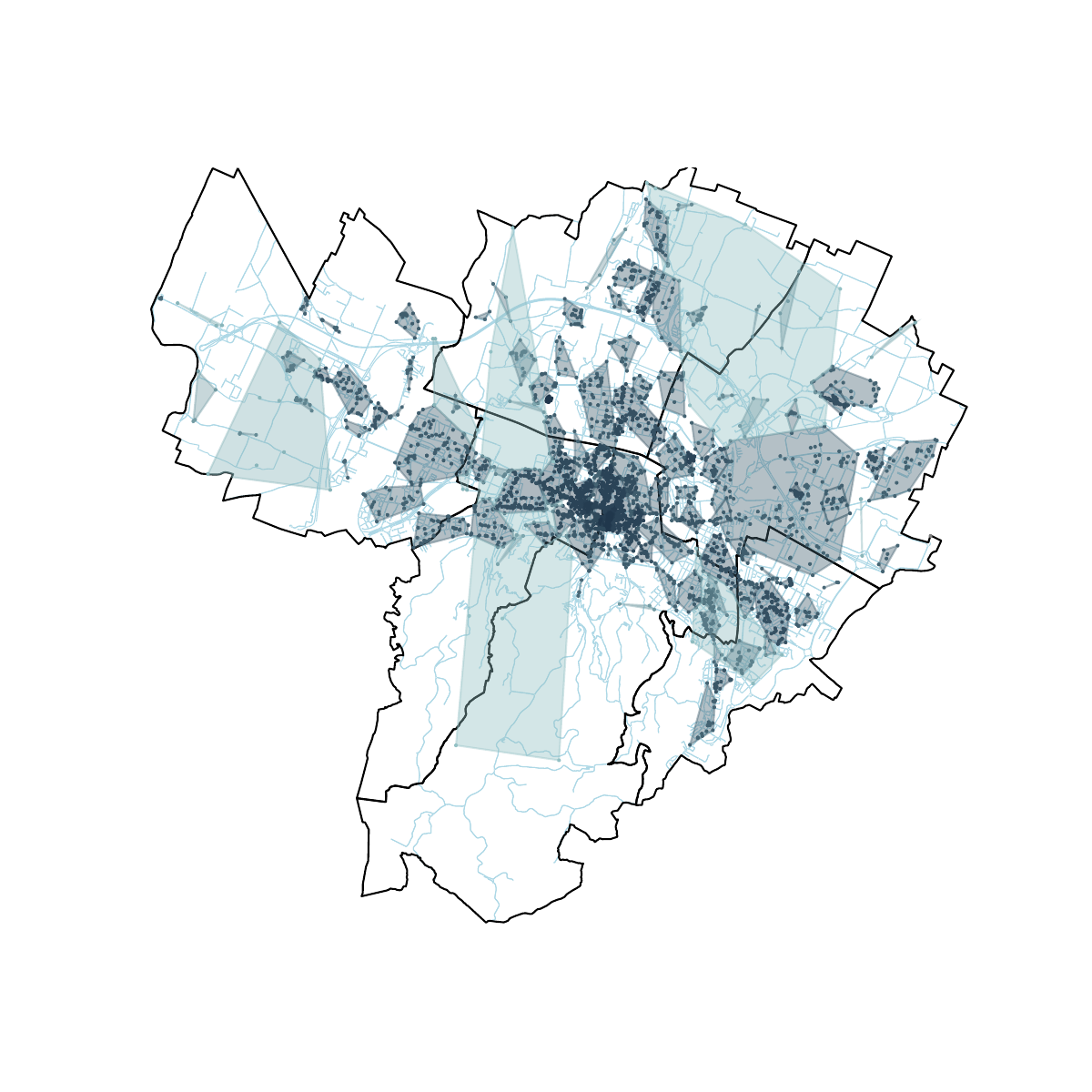}
    }\hspace{-1.3cm}
    \subfigure[{\shortstack{Travel and\\Transportation}}]{
      \includegraphics[width=.54\linewidth]{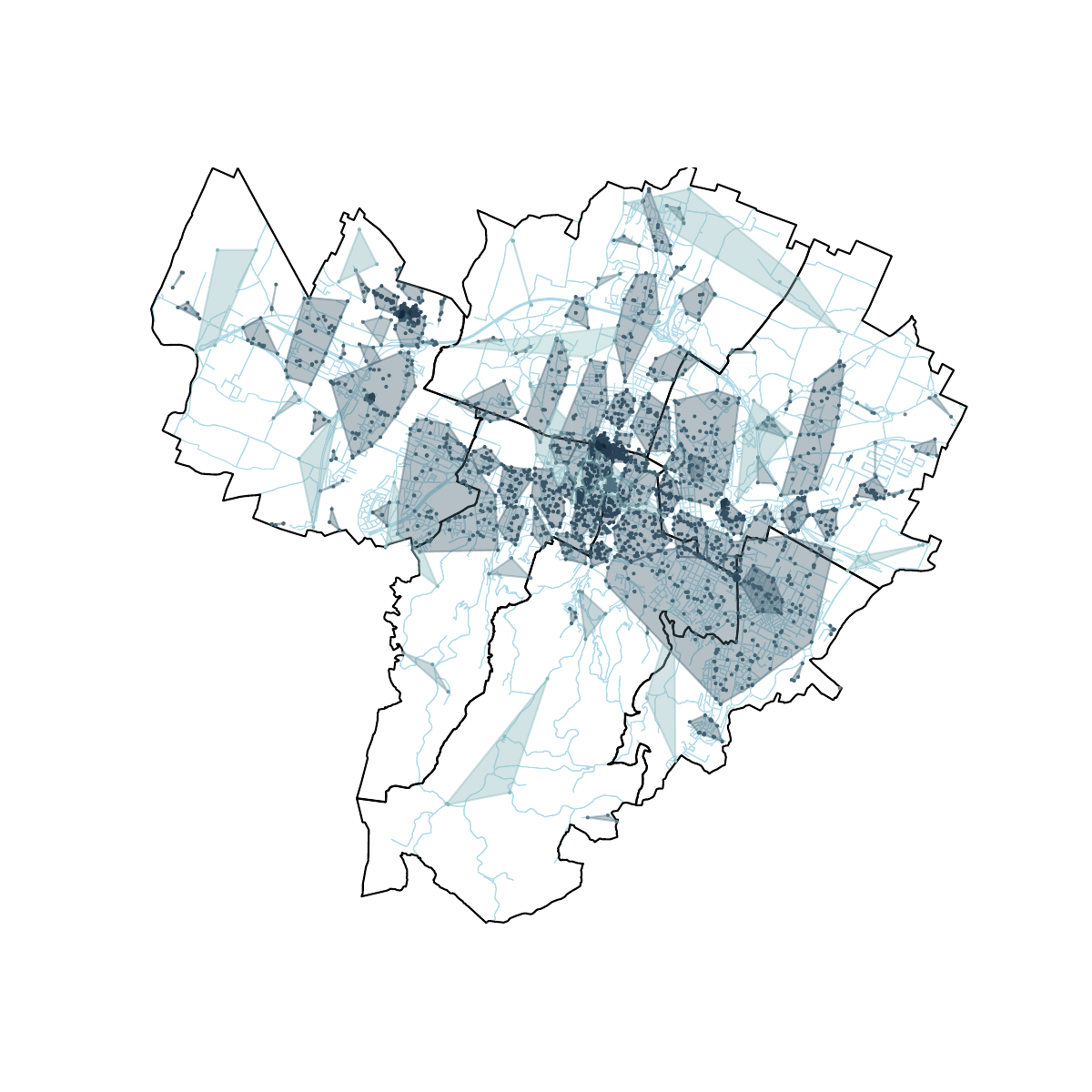}
    }
    \caption{Density-based clustering of POIs in Bologna, stratified by category. Categories show distinct spatial concentrations, with peak densities occurring in different parts of the city; grayscale encodes density (darker = higher), and not all categories span the full density range.}
    \label{fig:clustering}
\end{figure}

By systematically combining these parameter values, we generate~$30$ distinct DBSCAN-based clusterings of the POIs. Each clustering corresponds to a particular spatial and density resolution, enabling a multi-scale decomposition of the POI distribution. The resulting cluster configurations are shown in Fig.~\ref{fig:clustering}.

\begin{figure}[h!]
\centering
\subfigure[]{\includegraphics[width=.52\linewidth]{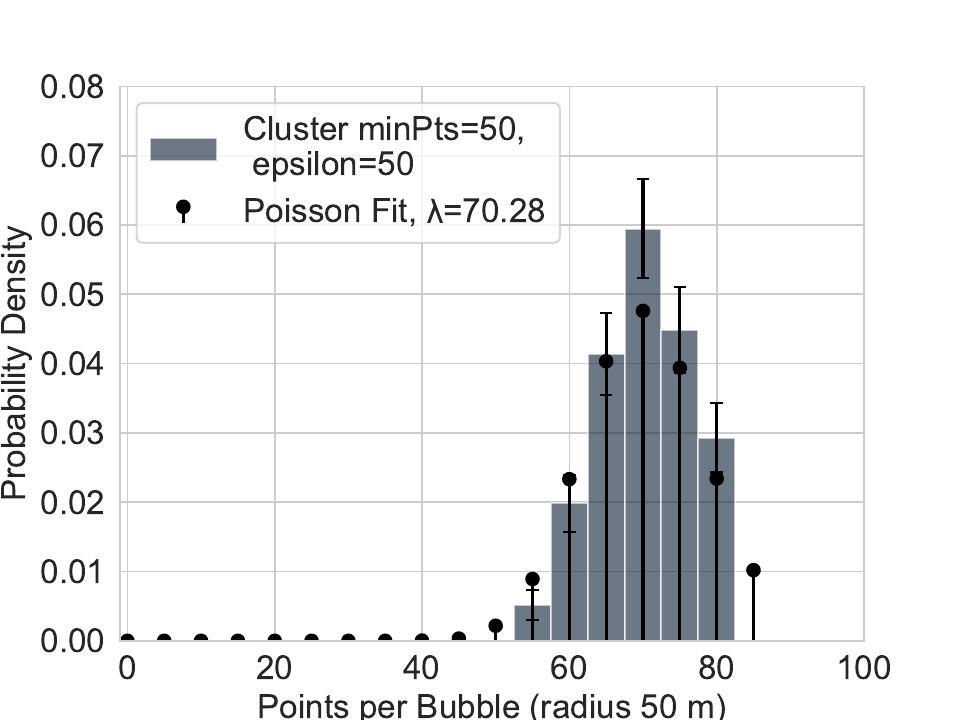}}\hspace{-.45cm}
\subfigure[]{\includegraphics[width=.52\linewidth]{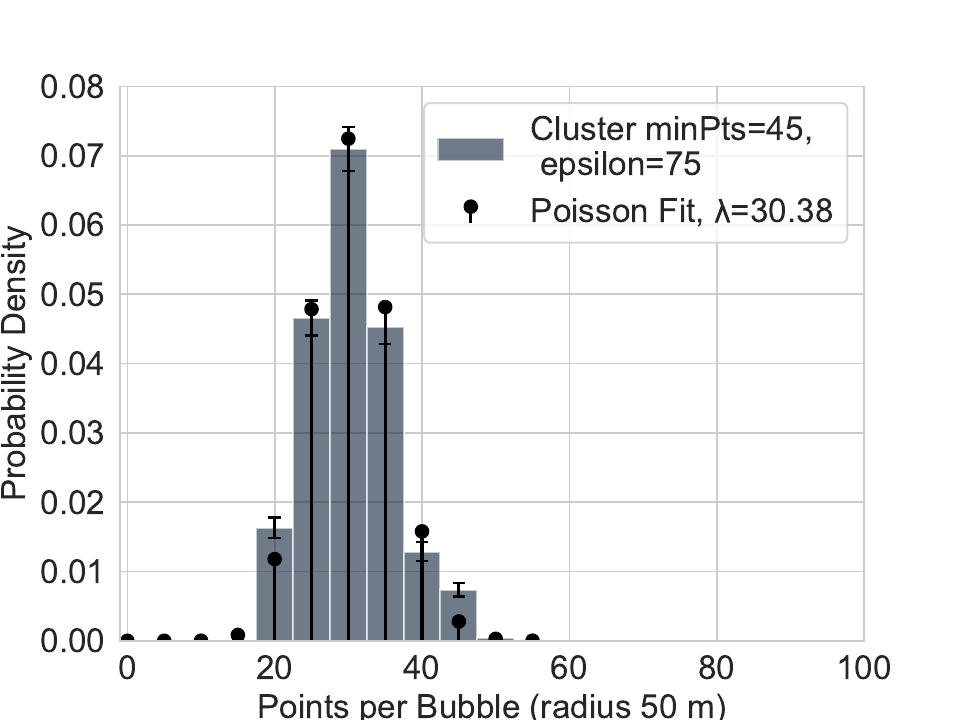}}
\subfigure[]{\includegraphics[width=.52\linewidth]{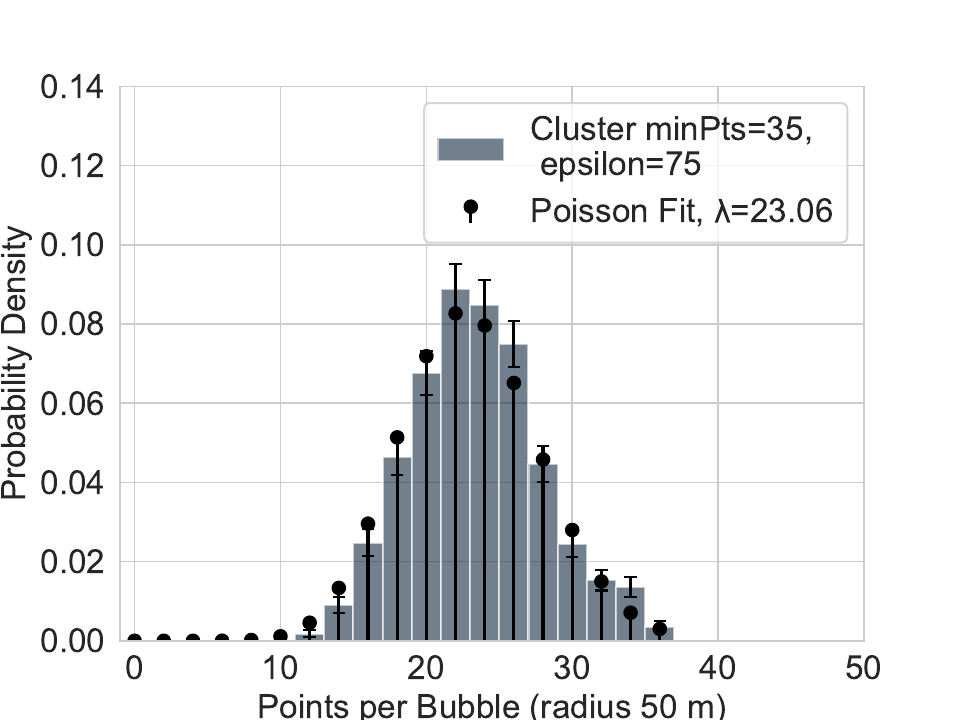}}\hspace{-.45cm}
\subfigure[]{\includegraphics[width=.52\linewidth]{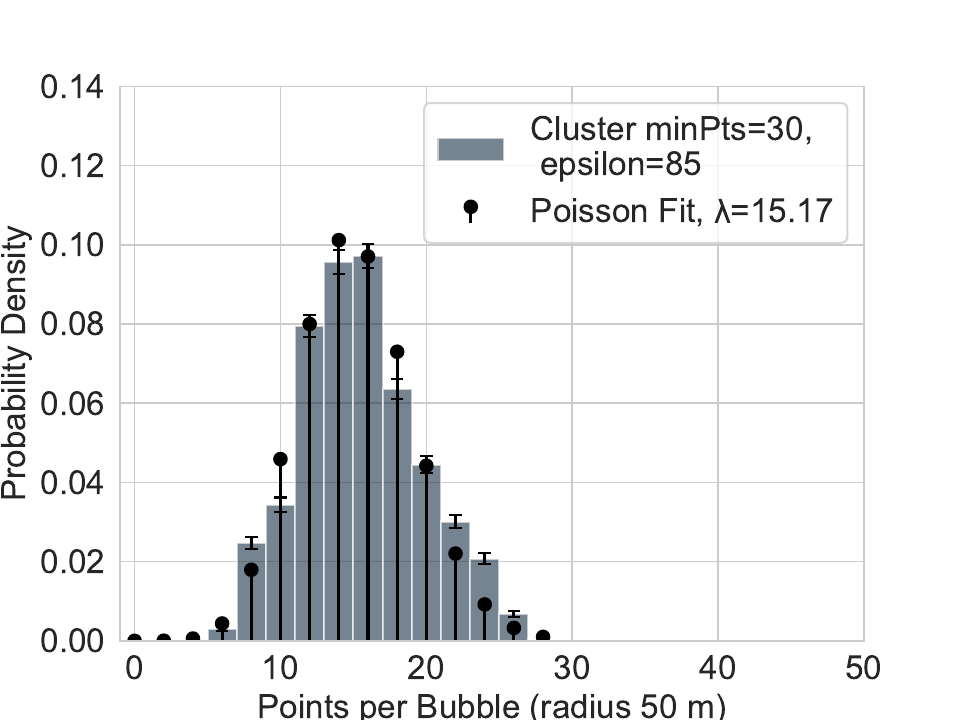}}
\subfigure[]{\includegraphics[width=.52\linewidth]{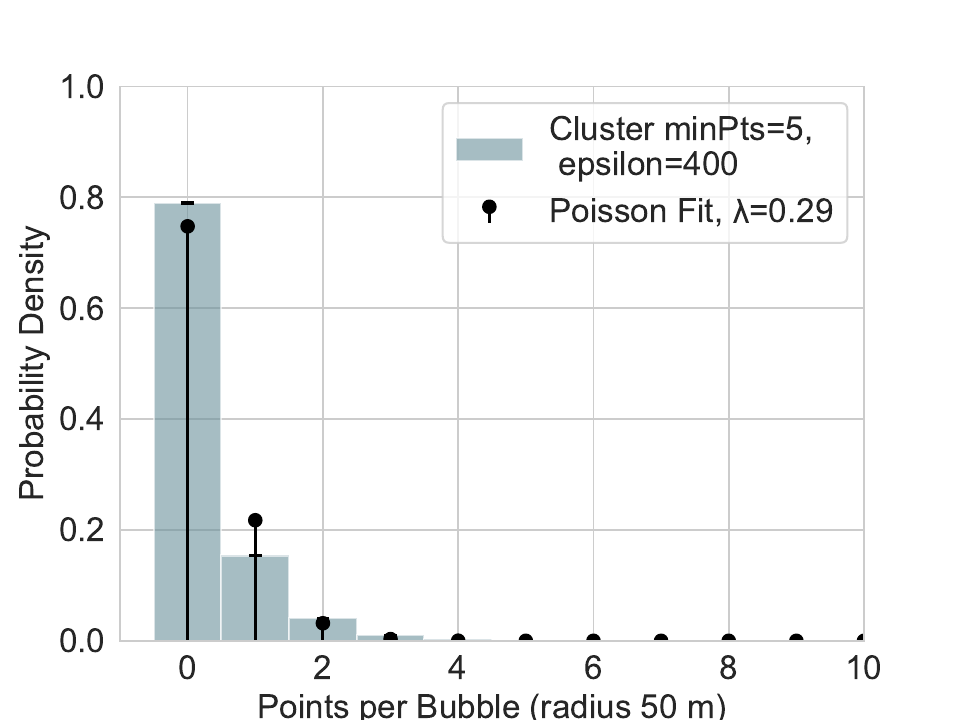}}\hspace{-.45cm}
\subfigure[]{\includegraphics[width=.52\linewidth]{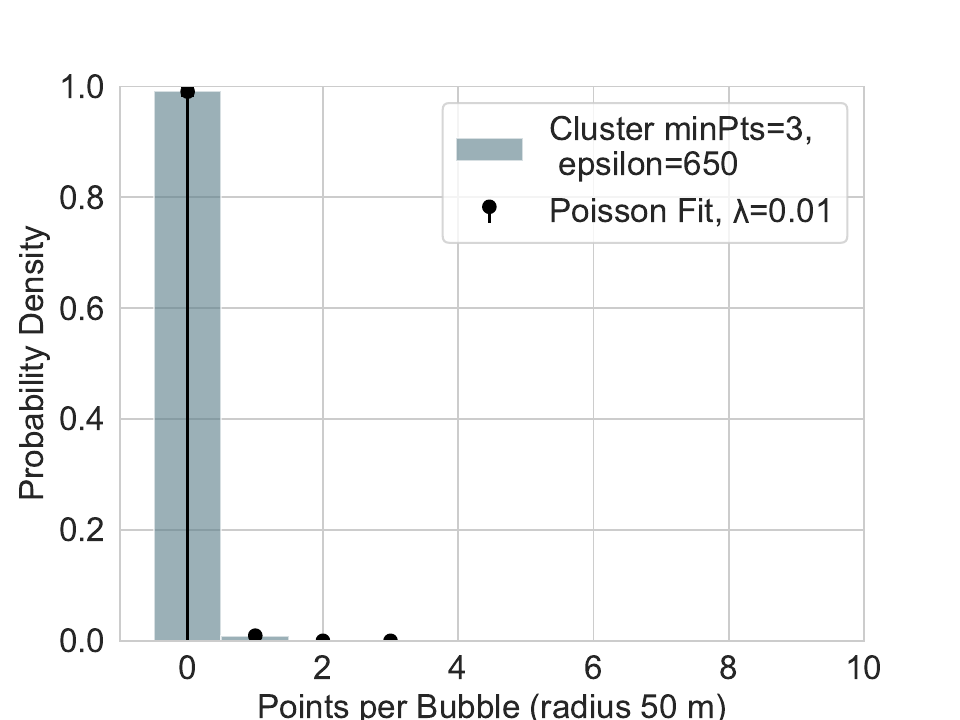}}
\caption{Histograms of POI counts per ball for selected clusters at different density thresholds, with Poisson fits. Panels (a-d) show the densest clusters, highlighting compact high-concentration regions; panels (e-f) depict more spatially diffuse clusters.}\label{fig:isto}
\end{figure}

This hierarchical process yields a layered decomposition of the urban surface into spatial clusters of varying densities, effectively isolating regions that are locally homogeneous in terms of POI concentration.\\
For each cluster obtained from the DBSCAN procedure, we compute the empirical distribution of POI counts across the balls contained within that cluster. These distributions, along with their corresponding error bars (representing the standard error of the mean in each bin), are then compared to the theoretical Poisson distribution with the rate parameter estimated from the sample mean. Fig.~\ref{fig:isto} shows a representative subset of these comparisons, selected from the 30 total clusterings obtained across all parameter combinations. In particular, the first four subfigures correspond to the densest clusters identified, highlighting compact regions with high POI concentration, while the last two represent more spatially diffuse clusters. While the Poisson fit generally provides a good approximation across all analyzed clusters, an exception is observed for the clustering obtained with \texttt{epsilon} = 50m and \texttt{minPts} = 50: in this case, the empirical distribution appears right-truncated around bin 90, with a noticeable overcount in the preceding bins (70–90). This pattern suggests the presence of a practical upper bound, possibly due to spatial constraints, on the number of POIs that can be accommodated within a 50m radius, leading to a pile-up effect just below the apparent saturation threshold.

The percentage of Bologna’s surface area covered by each identified cluster, stratified by POI category and clustering scale, reveal marked spatial heterogeneity in urban density patterns across functional domains.
Across all categories, a consistent pattern emerges: one or two large clusters, typically obtained with low-density thresholds (i.e., small \texttt{minPts} and large \texttt{epsilon}), dominate total surface coverage, reflecting spatially extensive, low-density regions that capture the widespread distribution of POIs. By contrast, a smaller number of clusters at higher density thresholds occupy much smaller portions of the urban surface, indicating compact areas with elevated local POI concentration. For the corresponding visualizations, see the Appendix (Figs.~\ref{fig:torte1}, \ref{fig:torte2}, \ref{fig:tortezoom1} and \ref{fig:tortezoom2}).\\
This distributional structure highlights a form of urban stratification: while many POIs are scattered diffusely across the urban landscape, certain categories give rise to highly localized, dense nuclei of activity. For instance, the \texttt{Retail} and \texttt{Travel and Transportation} categories show significant surface coverage by clusters formed at intermediate or high densities, suggesting the presence of well-defined commercial corridors or transit hubs. Conversely, the \texttt{Health and Medicine} and \texttt{Business and Professional Services} categories are almost entirely concentrated in low-density clusters, indicating a more homogeneous spatial distribution. These results underscore the utility of a multi-scale clustering approach in revealing the latent organization of urban functions and in differentiating between categories with distinct spatial footprints.

\section{Theoretical Framework: From Local Poisson Processes to Global Power Laws}\label{sec:math}

In this section, we provide a formal mathematical framework for modeling the spatial distribution of POIs within an urban environment. Our goal is to investigate how different generative assumptions at the local level give rise to distinct global patterns, specifically, the emergence of heavy-tailed distributions such as power-laws. We consider two complementary approaches: a spatially grounded model based on local Poisson processes, and a latent variable model based on Poisson mixtures.

The first model, presented in Section~\ref{sec:loc_poi}, assumes that POIs are locally distributed according to homogeneous Poisson processes within spatial clusters of varying density. In this setting, global heterogeneity emerges from spatial aggregation and variation in local intensities. The second model, introduced in Section~\ref{sec:mix_poi}, abandons explicit spatial structure in favor of a statistical mixture of latent components. Each observation is drawn from a Poisson distribution with an intensity sampled from a finite set of classes, capturing heterogeneity through probabilistic assignment rather than spatial partitioning. These two formulations allow us to contrast spatial and non-spatial mechanisms behind the emergence of heavy-tailed urban patterns. In Section~\ref{sec:hyb_poi}, we present a hybrid model that combines the strengths of both approaches, integrating spatial structure with latent components. This model allows us to simultaneously capture both spatial organization and the underlying probabilistic heterogeneity, offering a more flexible and comprehensive framework for understanding the emergence of heavy-tailed urban patterns. These three formulations enable us to contrast spatial and non-spatial mechanisms behind the emergence of such patterns.

\subsection{Spatial Heterogeneity via Local Poisson Processes}\label{sec:loc_poi}
In this section, we formalize the problem to provide a mathematical explanation for the global emergence of a power-law process, arising from locally Poissonian distributions \cite{10.1007/978-3-319-47886-9_15}.\\
Let \( S \) be a surface on which POIs are distributed. We define the random variable \( X_i \) as the number of POIs contained within ball \( i \), where each ball is a spatial unit of fixed radius (i.e., unit-area circles or tiles covering the surface, see Figure~\ref{fig:covering} in Appendix). In other words, the urban surface is partitioned into small, uniform-area spatial units, referred to as ``balls'', the term commonly used for them in Riemannian Geometry~\cite{jost2017riemannian}. For each ball \( i \), the number of POIs is denoted by an observed count \( X_i \in \mathbb{N} \).\\
The same surface \( S \) may also be regarded as being partitioned into a finite collection of larger clusters or regions (where we define a cluster as a union of overlapping balls, and a region as a union of clusters that are not necessarily contiguous), indexed by \( j = 1, \dots, C \), each with area \( A_j \). Within each region \( j \), the distribution of POIs in the constituent unit balls is assumed to follow a homogeneous Poisson process with intensity \( \lambda_j \). That is, the probability that a ball $i$ located in region \( j \) contains exactly \( k \) POIs is given by:
\[
P(X_i = k \mid A_j) = \frac{\lambda_j^k e^{-\lambda_j}}{k!}.
\]
To compute the total probability that a randomly chosen ball on the surface \( S \) has exactly \( k \) points, we need to weight the probabilities for each region by the relative area of the region. The probability of selecting a point in a given region \( j \) is proportional to the area \( A_j \) of that region relative to the total area \( A_S \) of the surface.\\
Thus, the total probability of having exactly \( k \) points is the weighted sum of the Poisson probabilities across all regions \( j \):
\begin{equation}\label{eq:1}
    P(X_i = k) = \frac{1}{A_S} \sum_{j=1}^{C} A_j \frac{(\lambda_j)^k e^{-\lambda_j}}{k!}
\end{equation}
where  \( C \) is the total number of regions, \( A_j \) is the area of region \( j \),
 \( \lambda_j \) is the Poisson intensity of the balls in the region \( j \),  \( k \) is the number of points, and \( A_S \) is the total area of the surface \( S \)
$$A_S=\sum_{j=1}^C A_j.$$
Equation~(\ref{eq:1}) provides the probability that a randomly chosen ball on the surface~\( S \) has {exactly~\( k \)} points in its unit neighborhood. This accounts for the varying Poisson intensities \( \lambda_j \) across different regions and the relative sizes of the regions.\\

\subsubsection{Local Poisson Distributions Leading to Global Power-Law Decay}

The goal is to show that local Poisson distributions can give rise to a global power-law decay.\\
Let $\lambda_j = j$ for $j \in \mathbb{N}$, corresponding to discrete Poisson components with increasing rate. This choice reflects the fact that the number of POIs per ball is integer-valued, making it natural to index the intensities by integers.\\
We now assume that the areas associated with each intensity level decay as a power law, i.e.,
\[
A_j \sim j^{-\beta}, \quad \text{with } \beta > 1.
\]
Under this assumption, the global distribution becomes
\[
P(X = k) \propto \sum_{j=1}^{C} j^{-\beta} \cdot \frac{j^k e^{-j}}{k!}.
\]
We analyze the asymptotic behavior of the series
\begin{equation}\label{eq:asymptotic_series}
    \sum_{j=1}^{C} \frac{j^{k - \beta} e^{-j}}{k!}
\end{equation}
in the joint limit \( C \to \infty \), \( k \to \infty \). Our objective is to characterize the rate at which the series tends to zero.\\
Applying Robbins’ bounds on the factorial~\cite{robbins1955remark}, we have
\[
k! = \sqrt{2\pi k} \left(\frac{k}{e}\right)^k e^{\theta_k}, \quad \theta_k \in \left[\frac{1}{12k+1}, \frac{1}{12k}\right].
\]
Accordingly, each term of the sum can be approximated by
\begin{equation}\label{eq:term_approx}
    \frac{j^{k - \beta} e^{-j}}{k!} \sim \frac{j^{k - \beta} e^{-j}}{\sqrt{2\pi k} \left(\frac{k}{e}\right)^k} = \frac{1}{\sqrt{2\pi k}} \left(\frac{j}{k}\right)^k j^{-\beta} e^{k - j}.
\end{equation}
This expression highlights the dependence on the ratio \( j/k \) and the exponential factor \( e^{k-j} \).\\

\paragraph*{Behavior for \( j \ll k \) and \( j \gg k \).}  
For \( j \ll k \), the term \( (j/k)^k \) tends to zero super-exponentially, while \( e^{k-j} \) grows exponentially. However, the decay from \( (j/k)^k \) dominates, causing these terms to vanish rapidly.\\
Similarly, for \( j \gg k \), the factor \( e^{k-j} \) decays exponentially, leading these terms to be negligible despite the growth of \( (j/k)^k \).
Therefore, terms outside a neighborhood of \( j \sim k \) contribute negligibly to the sum.\\

\paragraph*{Critical regime \( j \sim k \).}  
Let \( j = \beta k \) with \( \beta \in (0, \infty) \). Then the term in~\eqref{eq:term_approx} reads
\[
\frac{1}{\sqrt{2\pi k}} \beta^k (\beta k)^{-\beta} e^{k (1 - \beta)} = \frac{k^{-\beta}}{\sqrt{2\pi k}} \beta^{k - \beta} e^{k (1 - \beta)}.
\]
The dominant contribution arises near \( \beta = 1 \) (i.e., \( j \sim k \)), where the exponential factor \( e^{k(1-\beta)} \sim 1 \) and \( \beta^k \sim 1 \). At this peak, each term scales as
\[
\frac{j^{k - \beta} e^{-j}}{k!} \sim \frac{k^{-\beta}}{\sqrt{2\pi k}} = \mathcal{O}(k^{-\beta - 1/2}).
\]

\paragraph*{Total contribution near the peak.}  
The neighborhood \( j \in [k - \delta k,\, k + \delta k] \) for fixed \( \delta > 0 \) contains \( \Theta(k) \) terms. Summing over these terms yields
\[
\sum_{j \sim k} \frac{j^{k - \beta} e^{-j}}{k!} = \Theta(k) \cdot \mathcal{O}(k^{-\beta - 1/2}) = \mathcal{O}(k^{-\beta + 1/2}).
\]
Provided that \( C \to \infty \) sufficiently fast to include the critical region \( j \sim k \), the series~\eqref{eq:asymptotic_series} decays polynomially:
\[
\sum_{j=1}^C \frac{j^{k-\beta} e^{-j}}{k!} = \mathcal{O}(k^{-\beta + 1/2}).
\]
This power-law decay results from the interplay between the factorial growth in the denominator and the polynomial growth near the peak contribution.\\
In other words, the distribution of POI densities over a surface composed of regions whose areas scale proportionally to $j^{-\beta}$ and whose corresponding intensities of the Poisson distributions grow linearly with $j$ will exhibit an asymptotic behavior following a power law with a scaling parameter $\alpha := \beta - \tfrac{1}{2}$.


\textit{Remark.} It is worth emphasizing that the spatial contiguity of the regions is not a necessary condition for the emergence of global power-law behavior. In the mathematical formulation presented above, the global distribution arises from a weighted aggregation of local Poisson processes, where the weights are proportional to the surface areas $A_j$ and the associated intensities $\lambda_j$. As such, the key factor driving the emergence of heavy-tailed behavior lies in the joint distribution of areas and intensities, specifically, in the existence of appropriate scaling relationships between them. Whether the areas corresponding to different intensity levels are contiguous or spatially disjoint does not affect the resulting distribution, provided that their aggregate statistical contribution satisfies the required asymptotic conditions. This interpretation allows regions to consist of multiple spatially disjoint clusters, each composed of contiguous balls and sharing a common intensity. This enables the model to capture statistically homogeneous but spatially fragmented urban patterns.

\subsubsection{Three-Level Interpretation of the Local Poisson Model}
To summarize the modeling assumptions underlying the local Poisson framework, we introduce a structured three-level interpretation that connects the local generation of POIs with the emergence of global distributional patterns. This interpretation serves both to clarify the logic of the model and to enable direct comparison with the alternative mixture-based approach introduced in the next section.

\paragraph{Level 1 – Elementary spatial units (balls).}  
The urban surface \( S \) is partitioned into a set of uniform-area units, referred to as ``balls''. Each ball \( i \) is associated with an observed count \( X_i \in \mathbb{N} \), corresponding to the number of POIs located within it. At this level, no probabilistic model is assigned to individual balls: they represent the empirical substrate upon which higher-level structures are defined.

\paragraph{Level 2 – Homogeneous intensity regions.}  
Balls are first grouped into spatially contiguous clusters. Each cluster is a set of neighboring balls that form a connected area. Clusters that share the same Poisson intensity \( \lambda_c \) are then grouped into regions, indexed by \( c = 1, \dots, C \). 

Thus, each region \( c \) consists of one or more spatially disjoint clusters, all governed by the same homogeneous Poisson process. Within a given region, POI counts in individual balls are treated as independent and identically distributed:
\[
X_i \mid i \in c \sim \text{Poisson}(\lambda_c).
\]

The mesoscopic variability of POI density is therefore modeled through differences in intensity values across regions. This two-tiered structure, spatially contiguous clusters grouped into intensity-homogeneous regions, provides a flexible yet interpretable representation of urban space. It reflects the empirical observation that areas with similar POI density may be scattered across the city, while preserving the local spatial coherence within each cluster.

\paragraph{Level 3 – Global distribution and emergent power law.}  
The global distribution of POI counts across all balls is obtained by aggregating the contributions of all regions. If the distribution of region intensities \( \lambda_c \) follows a heavy-tailed law (e.g., a power law), then the resulting marginal distribution of \( X_i \) over the entire surface \( S \) also displays heavy tails. This mechanism provides a generative explanation for the emergence of a power-law regime in the empirical distribution of POI counts.


\subsection{Beyond Homogeneity: A Poisson Mixture Perspective}\label{sec:mix_poi}

The framework introduced in the previous section offers a mechanistic explanation for the emergence of global heavy-tailed distributions through the aggregation of spatially grouped Poisson processes with fixed intensities. While analytically tractable and conceptually insightful, this approach relies on strong structural assumptions, most notably, the existence of clearly defined regions characterized by homogeneous intensity levels and specific scaling relations between region size and rate. 

Such conditions may not fully hold in real-world urban environments, where transitions between areas are often gradual, boundaries are ill-defined, and local fluctuations in POI density can arise even within regions assumed to be internally homogeneous. Furthermore, enforcing a one-to-one mapping between spatial clusters and intensity values may overlook latent statistical variability that plays a key role in shaping global patterns.

To account for these limitations, we propose a model in which the local intensity \( \lambda_i \) is no longer fixed, but treated as a latent variable drawn from a finite set of possible values. This probabilistic formulation allows us to decouple spatial configuration from statistical structure, while preserving the Poissonian nature of local processes. Each ball is thus modeled as a realization of one of \( M \) latent intensity classes, each characterized by a Poisson rate \( \lambda_j \) and an associated mixing weight \( \pi_j \):

\[
X_i \sim \text{Poisson}(\lambda_j) \quad \text{with probability } \pi_j, \quad j = 1, \dots, M.
\]
As a result, the marginal distribution of POI counts over the city becomes a finite mixture of Poisson distributions:
\[
P(X_i = k) = \sum_{j=1}^{M} \pi_j \cdot \frac{\lambda_j^k e^{-\lambda_j}}{k!}.
\]

\subsection{Hybrid Model: Spatial Clustering with Intra-Region Poisson Mixtures}\label{sec:hyb_poi}
The Poisson mixture model presented in Subbsection~\ref{sec:mix_poi}, following \cite{Willmot2001}, can be viewed as a probabilistic analogy of the spatial model discussed in Section~\ref{sec:loc_poi}. In the spatial framework, global heavy-tailed distributions emerge from the aggregation of local Poisson processes associated with regions of varying surface area and intensity. The heterogeneity in region size and its assumed power-law scaling plays a central role in generating global scale-free behavior. By contrast, the mixture model dispenses with explicit spatial structure and instead introduces a latent class structure, where each ball is assigned to a Poisson component with probability $\pi_j$. Here, the role previously played by the physical area of a region is now assumed by the statistical weight of a latent class. In essence, the surface-driven heterogeneity that generates the power-law in the spatial model is replaced by class-driven heterogeneity in the mixture model.

This shift highlights a fundamental equivalence: in both frameworks, the emergence of power-law scaling is not inherently spatial but arises from the statistical distribution of Poisson intensities across contributing components, whether these components are spatially defined regions or latent classes. In this sense, the latent model captures the same generative mechanism, now abstracted away from geography and expressed through statistical mixing.

This conceptual bridge motivates the hybrid model introduced in Section~\ref{sec:hyb_poi}, which seeks to integrate these two perspectives. Rather than treating spatial and statistical heterogeneity as competing explanations, the hybrid approach defines latent intensity classes at the level of spatial regions—each viewed as an ensemble of clusters—thereby combining geographical structure with intra-regional variability. The result is a generative framework in which global heavy-tailed patterns emerge from the joint contribution of both surface heterogeneity and latent statistical mixing.

As a natural extension of the two models previously introduced, we now consider a hybrid hierarchical framework that incorporates both spatial organization and latent statistical heterogeneity. In this formulation, the urban surface is partitioned into regions made up of spatial clusters, as in the local Poisson model, but within each region, the distribution of POIs is governed by a finite Poisson mixture rather than a single homogeneous intensity.

Let \( S \) denote the total surface, divided into \( C \) regions indexed by \( c = 1, \dots, C \), each with area \( A_c \). Each region contains a set of unit-area spatial units (balls) indexed by \( i \in c \), where \( X_i \in \mathbb{N} \) denotes the observed POI count in ball \( i \).

Within each region \( c \), we assume that the distribution of \( X_i \) follows a mixture of \( M_c \) Poisson components. Each component is characterized by an intensity \( \lambda_{cj} \) and a corresponding weight \( \pi_{cj} \), with:
\[
\sum_{j=1}^{M_c} \pi_{cj} = 1.
\]
The intra-region distribution of POI counts is thus given by:
\begin{equation}
P(X_i = k \mid i \in c) = \sum_{j=1}^{M_c} \pi_{cj} \cdot \frac{\lambda_{cj}^k e^{-\lambda_{cj}}}{k!}.
\label{eq:cluster_mixture}
\end{equation}

Aggregating across all regions and weighting each region according to its relative area, the global distribution of POI counts over the entire surface \( S \) becomes:
\begin{equation}
P(X = k) = \sum_{c=1}^{C} \frac{A_c}{A_S} \cdot \left( \sum_{j=1}^{M_c} \pi_{cj} \cdot \frac{\lambda_{cj}^k e^{-\lambda_{cj}}}{k!} \right),
\label{eq:global_mixture}
\end{equation}
where \( A_S = \sum_{c=1}^{C} A_c \) is the total area.

This model generalizes both the local Poisson and the Poisson mixture frameworks. When \( M_c = 1 \) for all \( c \), Equation~\eqref{eq:cluster_mixture} reduces to the local Poisson model. Conversely, if \( C = 1 \), i.e., the entire surface is treated as a single region, Equation~\eqref{eq:global_mixture} recovers the global Poisson mixture model. The hierarchical formulation thus provides a flexible structure capable of capturing both spatially structured and statistically heterogeneous patterns in urban POI distributions.

\subsection*{Distributed Contributions to Power-Law Tails}

In contrast to the earlier models, the hierarchical mixture formulation allows power-law behavior to emerge as a collective effect of both spatial and statistical structure. Let \( w_{cj} = \frac{A_c}{A_S} \cdot \pi_{cj} \) denote the weight associated with the Poisson component of intensity \( \lambda_{cj} \). Then, the total contribution to the tail at level \( k \) is approximately governed by:
\[
\sum_{(c,j):\ \lambda_{cj} \sim k} w_{cj} = \sum_{(c,j):\ \lambda_{cj} \sim k} \frac{A_c}{A_S} \cdot \pi_{cj}.
\]
If this aggregate weight satisfies the scaling relation
\[
\sum_{(c,j):\ \lambda_{cj} \sim k} \frac{A_c}{A_S} \cdot \pi_{cj} \sim k^{-\beta},
\]
then, using the asymptotic estimate derived previously for a Poisson component with intensity \( \lambda \sim k \), each term contributes \( \mathcal{O}(k^{-\beta - 1/2}) \) to the total. Hence, the global distribution satisfies
\[
P(X = k) \sim k^{-\beta - 1/2}
\quad \text{as } k \to \infty.
\]

This condition allows multiple clusters of different sizes to jointly contribute to a common intensity level. For example, two clusters \( c_1 \) and \( c_2 \) may each contain components with intensity \( \lambda_{cj} \sim k \), and although neither cluster alone exhibits a power-law scaling in area, the combined contribution of their surface-area-weighted mixture probabilities can still decay as \( k^{-\beta} \), producing a global power-law tail. The emergence of the tail is thus not due to the structure of individual clusters, but rather to the statistical geometry of the entire ensemble. This highlights the hierarchical model’s capacity to encode and synthesize heterogeneous local behaviors into coherent macroscopic patterns.

\section{Empirical Validation of the Poisson Mechanism}\label{sec:val}
After proposing a mathematical formalism for the emergence of heavy-tailed via the superposition of locally homogeneous processes, we verify its viability through computer simulations. \\
To validate the theoretical framework outlined in Section \ref{sec:math}, we now turn to an empirical assessment of its generative capabilities. In particular, we aim to test whether the Hybrid Poisson mixture model introduced in Section \ref{sec:hyb_poi} - parameterized to reflect the observed heterogeneity in urban POI distributions - can reproduce the heavy-tailed behavior and power-law characteristics identified in the empirical data. 
This step is intended to strengthen the connection between the theoretical model and empirical observations. While the previous section demonstrated, in principle, how heavy-tailed distributions can emerge from aggregated Poissonian processes, we now explore whether this mechanism also aligns with the empirical data. Specifically, we calibrate the hybrid Poisson mixture model using observed POI counts and assess its ability to reproduce the key statistical features of the real distribution, including the presence of a power-law regime. 
For this purpose, we first construct a synthetic surface capable of simulating the theoretical results demonstrated in Section~\ref{sec:loc_poi}, which we will empirically validate as a starting point. Subsequently, we define the latent probabilities introduced in Section ~\ref{sec:hyb_poi}, in such a way that they allow us to describe the real model through the theoretical results demonstrated.

\subsection{Synthetic Construction of Scale-Free Composite Surfaces via Local Poissonian Processes}
\begin{figure}[htp]
    \centering
    \includegraphics[width=1\linewidth]{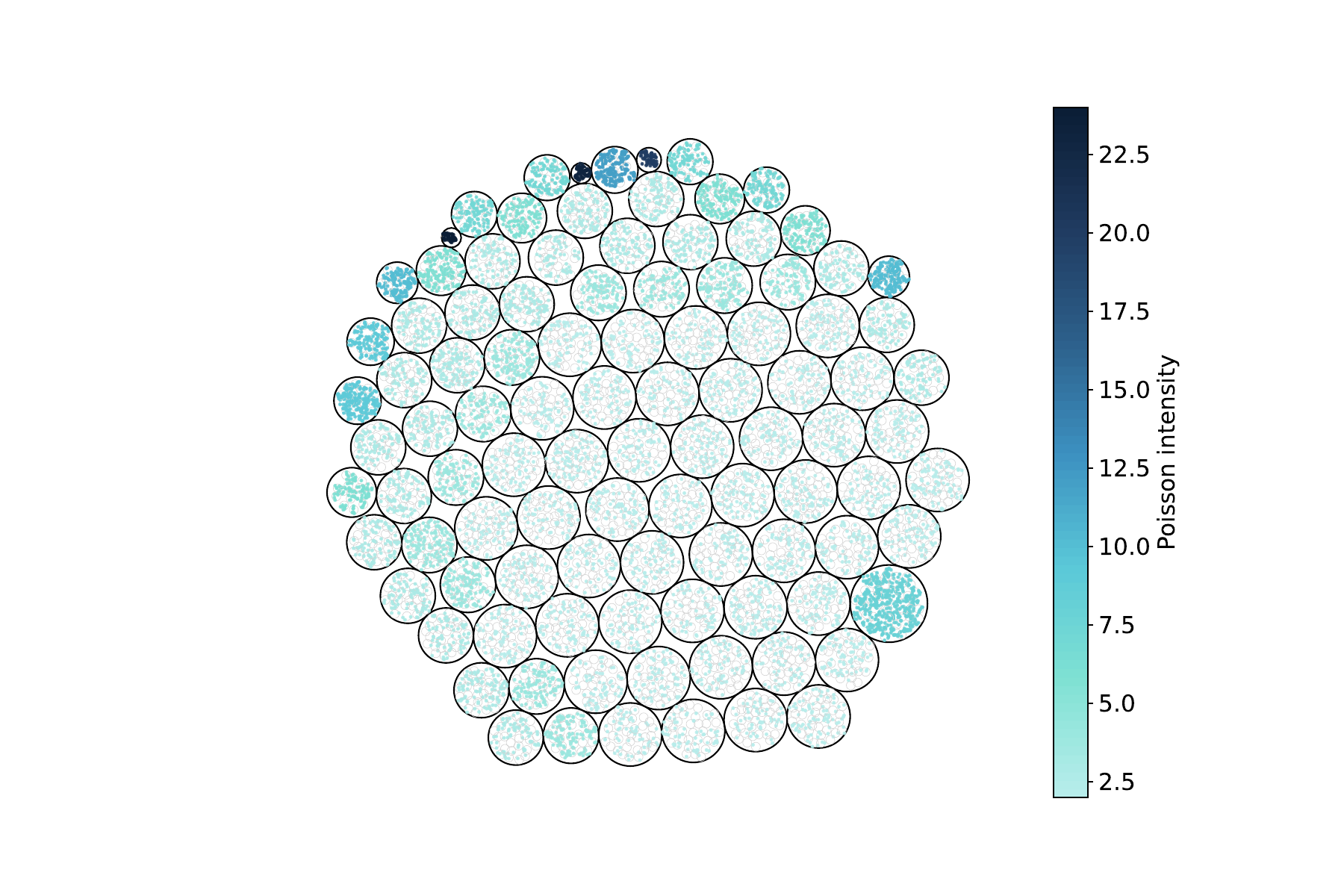}
    \caption{
Synthetic composite surface generated by partitioning the plane into circular regions, represented in black, whose areas scale as a power law of the assigned intensity class with exponent $\beta=2.5$. Each region is tessellated by unit-area balls, within which finer-scale local Poisson point processes are sampled.
}\label{fig:sinteticsurface}
\end{figure}

\begin{figure}[htp]
    \centering
    \includegraphics[width=0.95\linewidth]{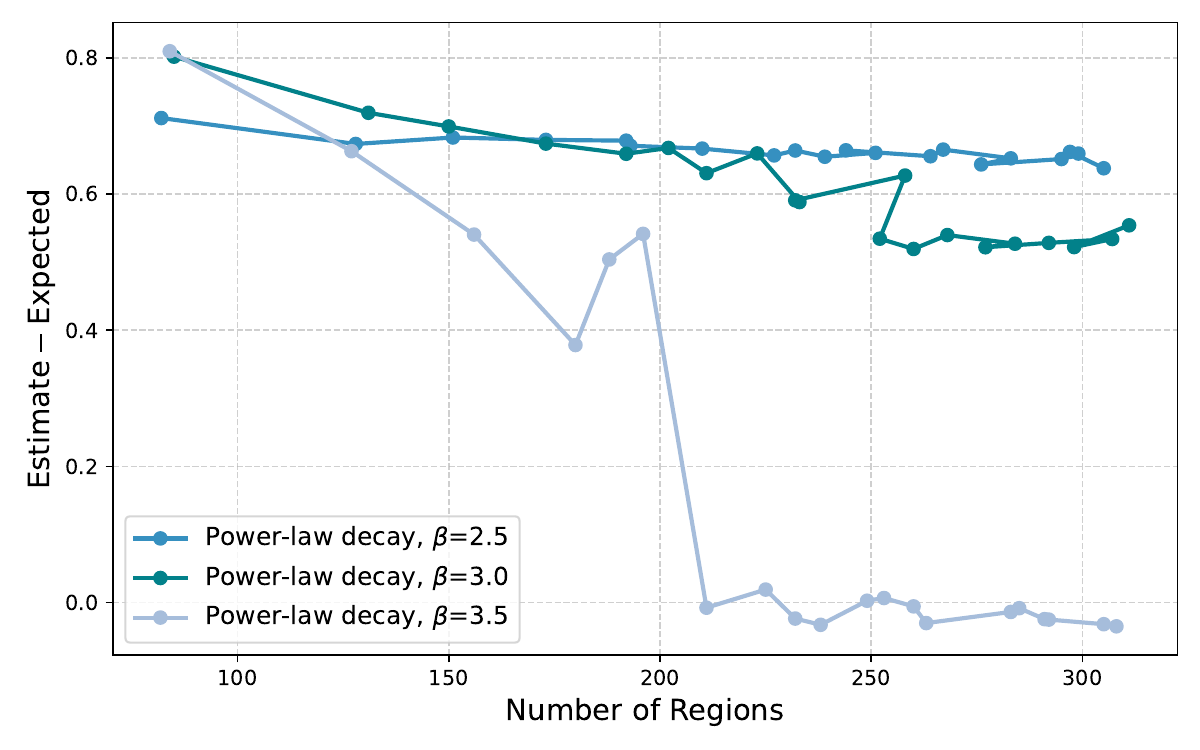}
\caption{Error between estimated and expected decay exponents (for $\beta=2.5$, $\beta=3.0$ and $\beta=3.5$) across varying numbers of regions.
}\label{fig:alphadecay}
\end{figure}

Following the hierarchical framework introduced in Section~\ref{sec:loc_poi}, we construct a synthetic model in which local Poisson point processes are embedded into spatial units whose sizes follow a power-law distribution.\\
Inspired by the Barabási–Albert preferential attachment model~\cite{barabasi1999emergence}, we construct a synthetic spatial surface characterized by a heavy-tailed distribution of the areas of its regions.
In our construction, the \textit{nodes} represent clusters, and nodes sharing the same degree represent regions.\\
Namely, for each node of degree $j$, we define a \emph{cluster}, that is, a circular surface whose area scales as:

$$A_j = \frac{j^{-\beta}}{n_{j}},$$
where $n_{j}$ denotes the number of clusters having intensity $j$, and $\beta > 1$ is a tunable exponent controlling the decay. This normalization ensures that the total area associated to all clusters of the same intensity, i.e. each region, remains consistent with the scaling assumption:
\[
\sum_{j:\,j=\lambda_j} A_j = j^{-\beta}.
\]
As a result, the model yields many regions of varying size, where higher-intensity regions are smaller on average, consistent with empirical observations of scale-free heterogeneity.\\
As in Section~\ref{sec:loc_poi}, where regions are defined as conceptual aggregates of all spatial units sharing the same intensity, the union of all clusters with identical intensity in our construction is, in aggregate, equivalent to the corresponding theoretical region. This correspondence ensures that the total area associated with each intensity class respects the scaling law, while still preserving a granular spatial representation based on distinct clusters.\\
Each cluster is then densely tessellated by \emph{balls}---unit-area disks that serve as elementary sampling units. Within each ball, a Poisson point process is independently sampled with intensity $\lambda_i = i$, inherited from the parent cluster.\\
Importantly, all local and global statistical distributions are computed over the set of balls. Therefore, the distinction between clusters and aggregated regions does not affect the measurement of the point count distributions, which depend only on the intensities and the arrangement of the balls rather than on whether clusters are grouped into larger regions.\\
Fig.~\ref{fig:sinteticsurface} shows a synthetic surface generated through this procedure. Each disk represents a cluster, and disks sharing the same color represent regions. 
\begin{figure}[htp]
    \centering
    \includegraphics[width=0.9\linewidth]{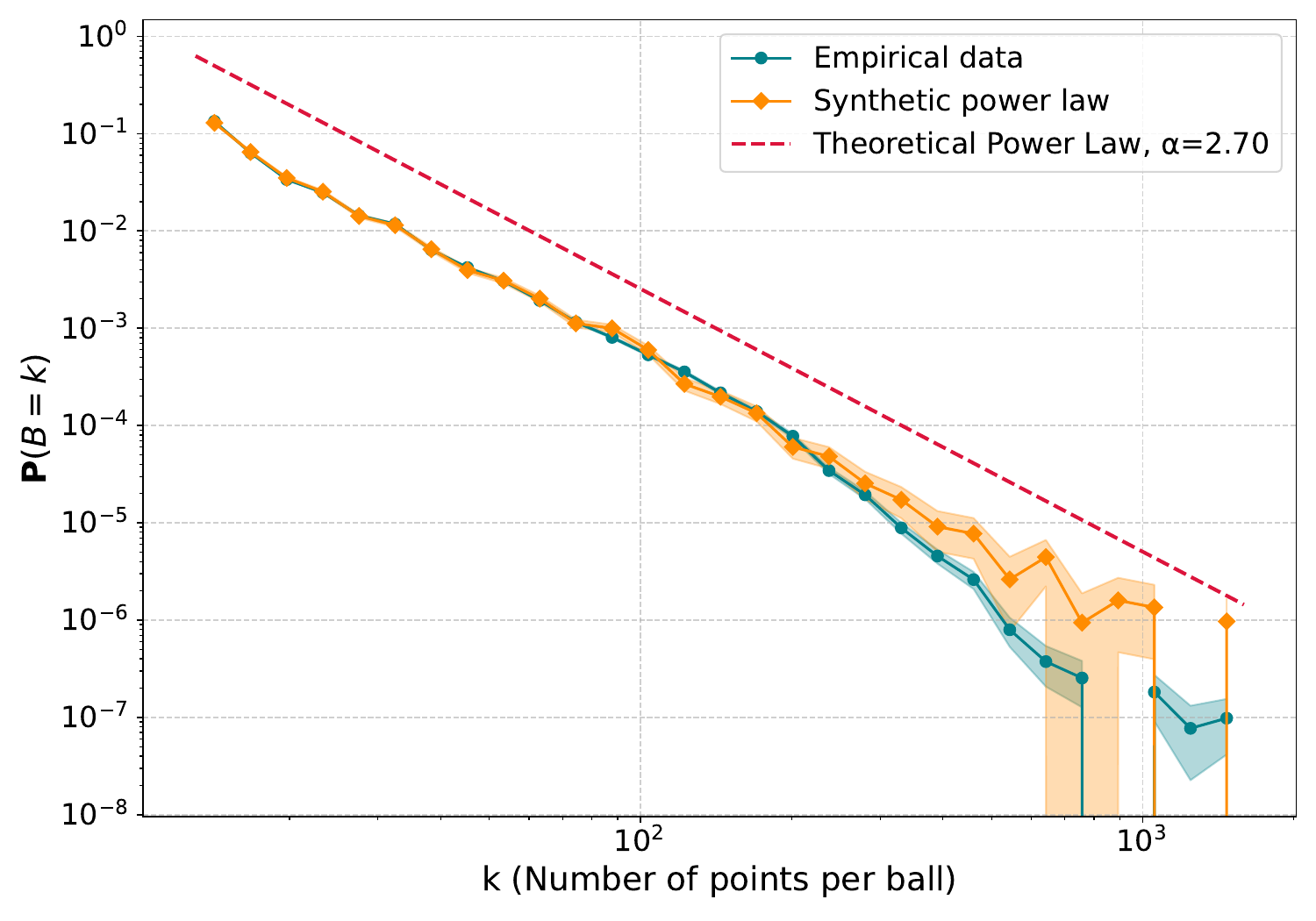}
\caption{
Distribution of the number of points per unit ball in the synthetic surface generated with $\beta=2.5$, $10^6$ clusters organized into $\approx 300$ regions and unit ball area $3 \times 10^{-7}$. The resulting distribution exhibits a power-law decay with exponent $\alpha \approx 2.7$.
}\label{fig:sinteticlog}
\end{figure}
Fig.~\ref{fig:sinteticlog} displays the resulting distribution of the number of points in unit balls in the case of a surface generated with $\beta = 2.5$, $10^6$ clusters organized into approximately 300 regions, and a unit ball area of $3 \times 10^{-7}$. In Section~\ref{sec:loc_poi}, we demonstrated that for regions whose area scales as a power law $j^{-\beta}$ and whose corresponding Poisson intensities scale linearly with $j$, the resulting distribution exhibits a power-law decay. Here, we support this result by showing that the empirical distribution does not significantly deviate from a power-law model, according to the KS test (D = 0.0123, p-value = 0.0625).\\
The exponent of the fitted power law is~$\approx 2.7$. Although the theoretical exponent expected from the mathematical derivation is $\alpha=\beta - 1/2 = 2$, the observed discrepancy is due to the finite size of the simulations. In fact, as the number of clusters increases, the estimated exponent progressively approaches the expected value, as shown in Fig.~\ref{fig:alphadecay}. Interestingly, we observe that larger values of $\beta$ lead to faster convergence toward the asymptotic decay. This behavior can be explained by the properties of the area distribution's tail: when $\beta$ is small, the probability of generating very large regions is higher, resulting in heavier tails and a greater variance in the number of points per region. As a consequence, much larger samples are required to stabilize the empirical distribution and recover the theoretical scaling. Conversely, when $\beta$ increases, the distribution of region sizes becomes more concentrated, extreme values are less frequent, and the variance decreases. This leads to a more rapid convergence to the predicted power-law behavior, even with a smaller number of clusters. This observation underscores an important methodological consideration: the rate at which empirical estimates approach theoretical expectations depends strongly on the tail heaviness of the underlying distributions.

\subsection{Verification of the theoretical conditions}
\begin{figure}[h!]
    \centering
    \subfigure[]{\includegraphics[width=.52\linewidth]{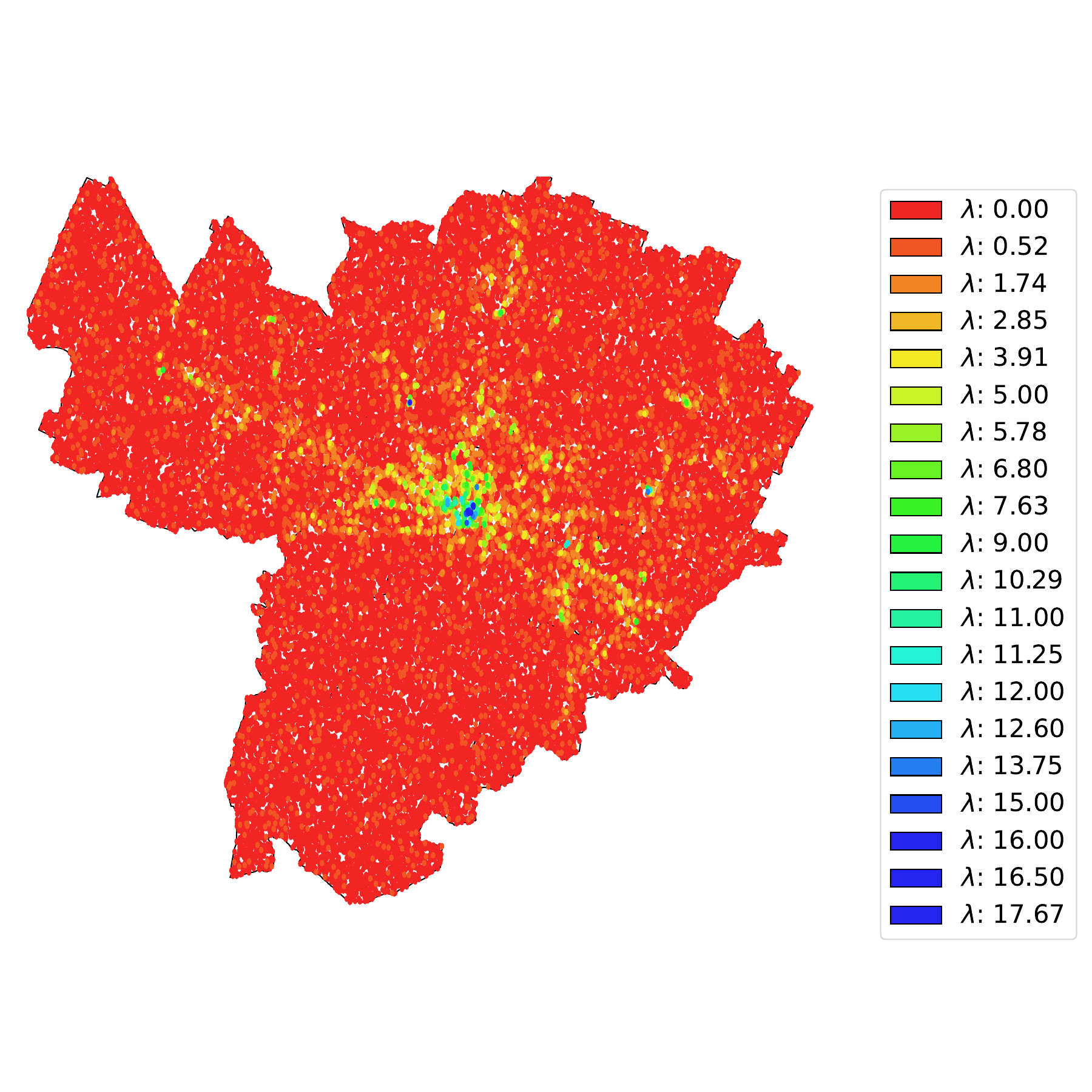}}\hspace{-.1cm}
    \subfigure[]{\includegraphics[width=.47\linewidth]{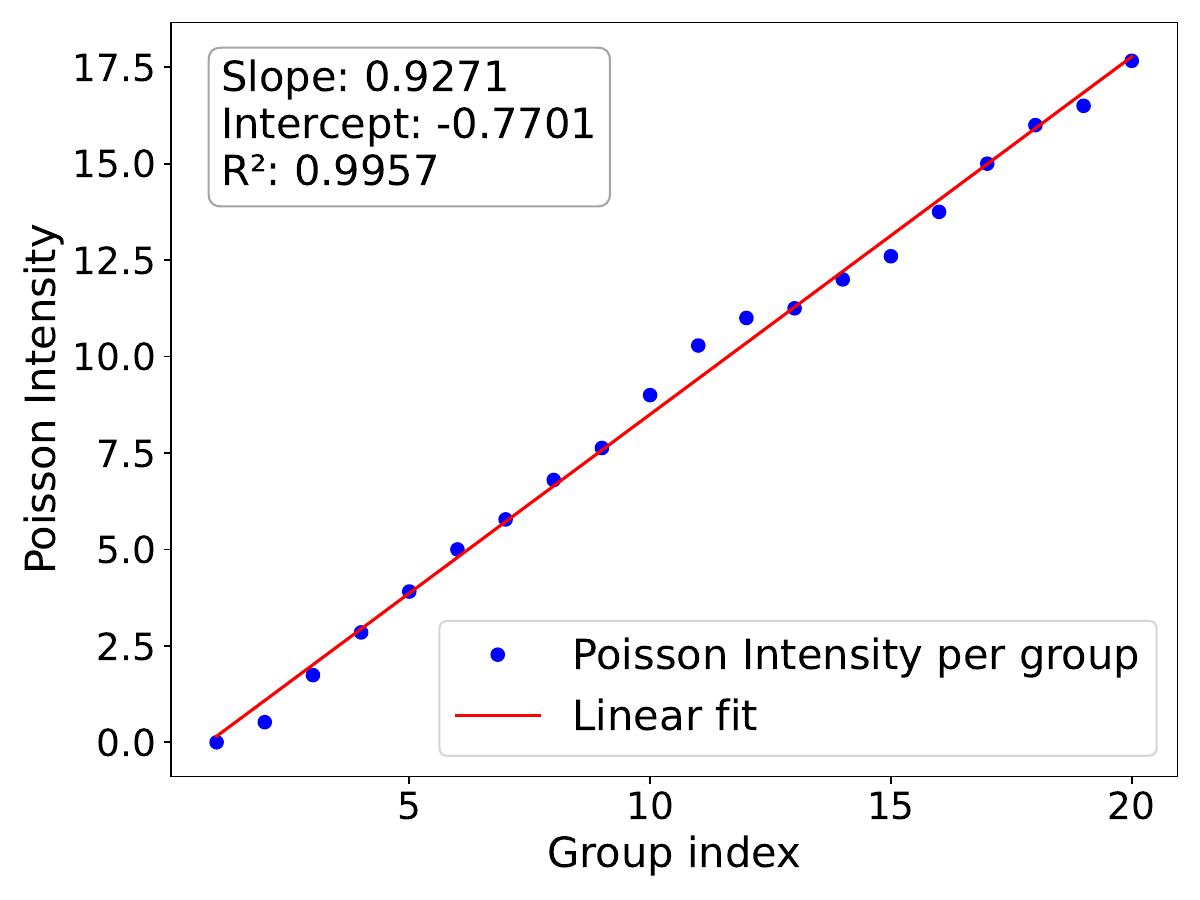}}
    
    \textbf{Retail}
    
    \vspace{0.8em}
    
    \subfigure[]{\includegraphics[width=.52\linewidth]{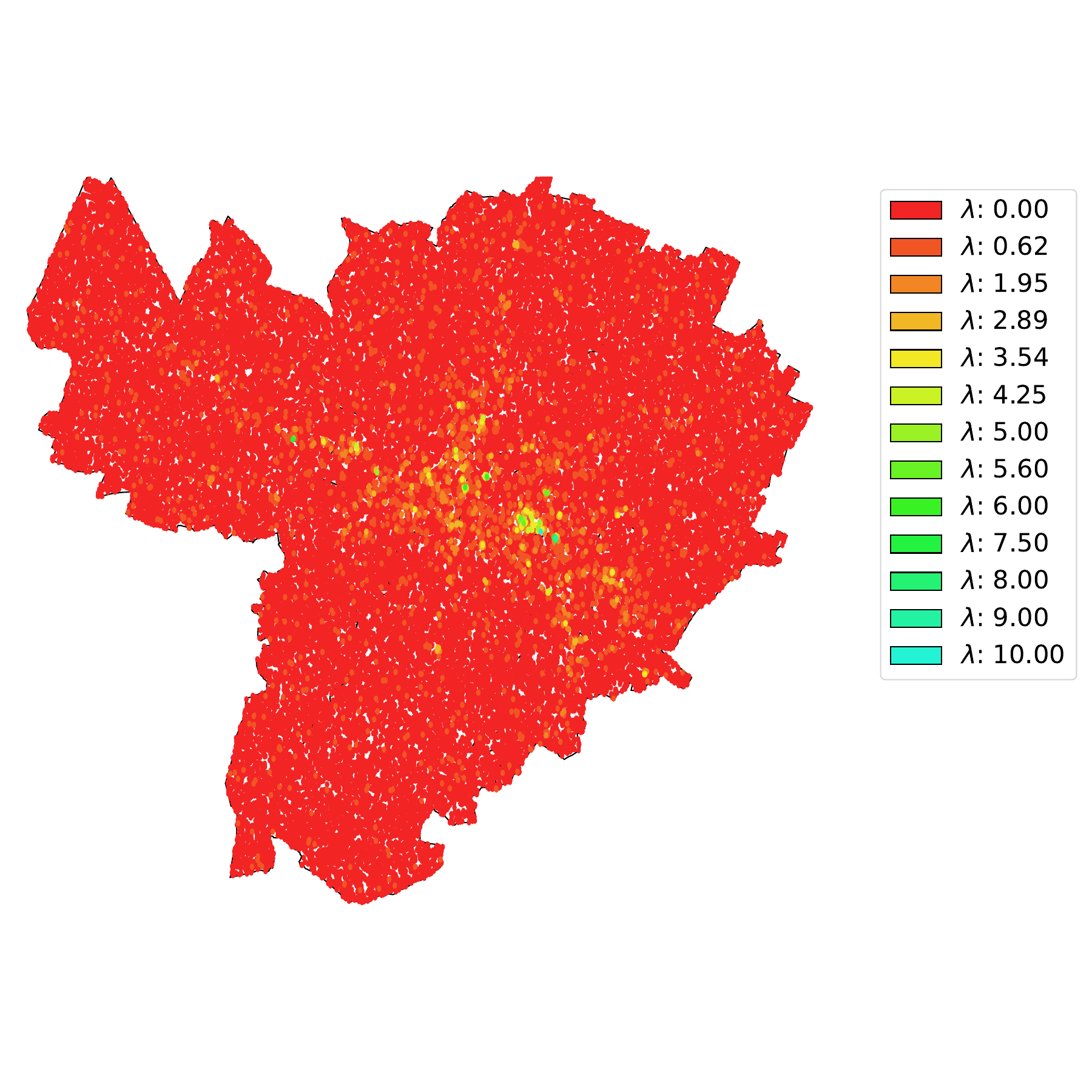}}\hspace{-.1cm}
    \subfigure[]{\includegraphics[width=.47\linewidth]{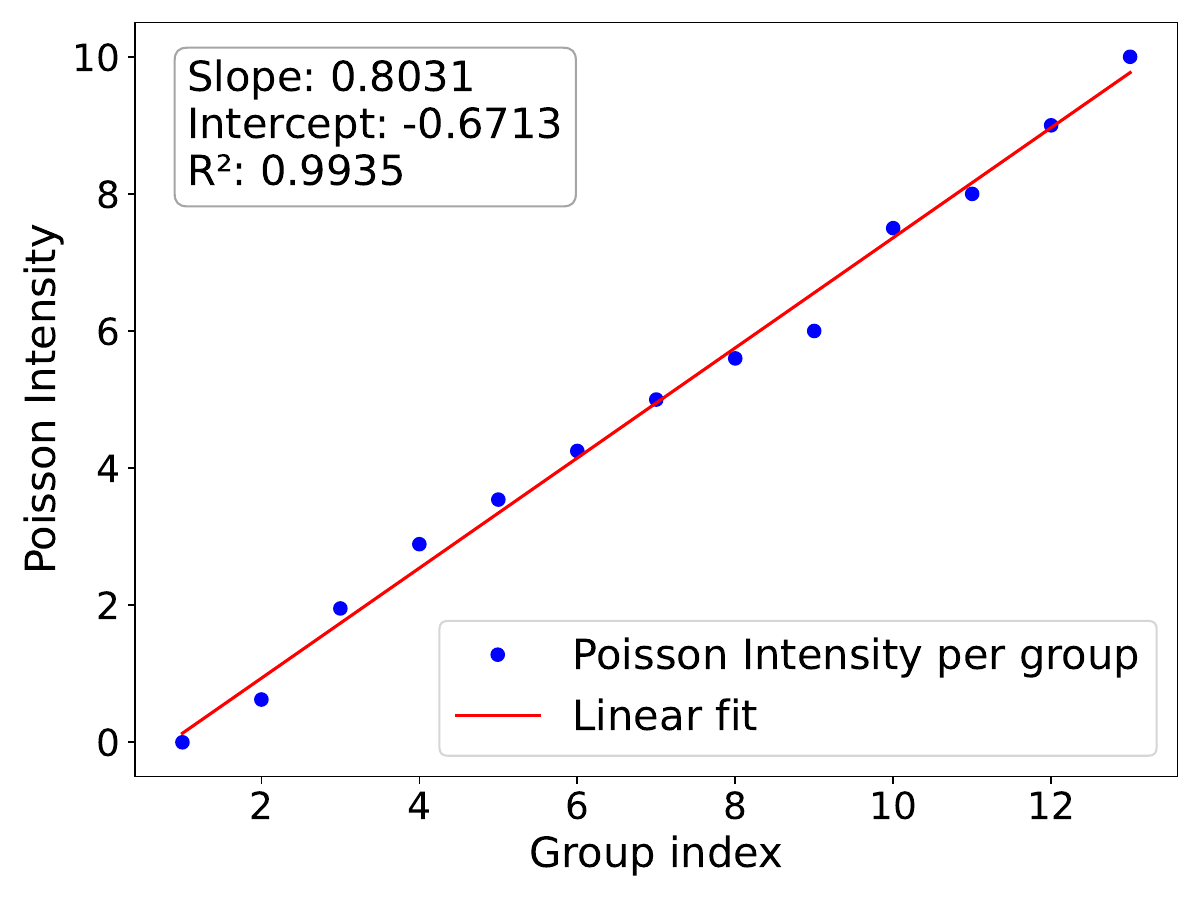}}
    
    \textbf{Health and Medicine}
    
    \caption{Left: Unitary balls on the map colored by regions, with region sizes following a power law scaling based on their cardinality. Right: Linear fit of POI distribution intensities within unitary balls per region, highlighting the relationship between area size and point density.}\label{fig:gruppi}
\end{figure}

In this section, we aim to verify whether our empirical data satisfy the conditions under which the results demonstrated in Section~\ref{sec:loc_poi} hold exactly. 
In Section~\ref{sec:met}, we showed that the real-world POI data from the city of Bologna can be globally described by a power-law distribution and locally by Poisson distributions.
However, the relationships linking the areas and intensities derived from the data appear to be quite distant from those presented in Section~\ref{sec:loc_poi}, which requires strong assumptions to be valid.\\
Here, we demonstrate that by introducing latent probabilities and employing the hybrid model, we can bridge these two interpretations. To this end, we re-cluster the unitary balls into new regions. Specifically, we first compute the scaling parameter $\alpha$ of the global distribution for the category and predefine the number of regions.
Each region is then composed of a number of unitary balls proportional to the region index raised to the power of $-\bigl(\alpha + \tfrac{1}{2}\bigr)$. In this way, we partition the surface into areas whose sizes scale according to a power law.\\
We sort the unitary balls in ascending order according to the number of POIs contained in each, and assign them sequentially to the regions. Fig.~\ref{fig:gruppi} shows the result of this procedure for the \texttt{Retail} and \texttt{Health and Medicine} categories, and illustrates how the Poissonian intensity measured in each region grows linearly with the region index, as it is expected from our model. This provides striking empirical validation of the theoretical result. Indeed, we recover all the assumptions required for the generation of a power-law distribution with exponent $\alpha$, namely regions characterized by Poissonian distributions whose intensities scale linearly with the region index (see Fig.~\ref{fig:gruppi}(b) and (d)), and region areas that follow a power law with exponent $\beta$.
Moreover, we observe that categories characterized by higher values of $\alpha$ require fewer regions to adequately capture the overall distribution of points, as the tail of the distribution decays more rapidly.\\
Already this alternative sampling of the regions confirms that it is indeed possible to explain the emergence of the power-law distribution purely in terms of local Poisson processes. However, although this procedure theoretically clarifies the link between the two distributions, it has limited practical significance, since the elementary balls forming each region can be potentially very distant or spatially isolated from each other.\\
By instead redefining the regions introduced in Section~\ref{sec:met} through latent probabilities associated with corresponding Poisson intensities, which are simply the proportions of elementary balls identified in the new clustering, we obtain an exact correspondence between the hybrid model and reality.\\
In practice, we retain the hierarchical clustering, which yields a small number of representative regions. These regions consist of clusters (i.e., unions of contiguous unitary balls) that share local Poissonian point distributions, while simultaneously being describable through latent probabilities that account for the emergence of the global power-law behavior.

\section{Conclusion}\label{sec:con}

In this work, we proposed a novel theoretical and empirical framework to explain how power-law distributions in the spatial organization of urban POIs can emerge purely from the aggregation of locally homogeneous Poissonian processes. Unlike explanations rooted in criticality, self-organization, or other endogenous complex mechanisms, our results demonstrate that statistical heterogeneity in the distribution of local intensities is sufficient to generate heavy-tailed behavior at the macroscopic scale.\\
By formally deriving the scaling conditions under which this mechanism produces power-law distributions, we contribute a rigorous analytical foundation that links micro-level spatial regularity with emergent global complexity. Beyond its theoretical significance, this approach offers a new perspective for the analysis and classification of urban space. The combination of hierarchical clustering and latent Poisson mixtures provides a flexible modeling strategy capable of capturing both spatial contiguity and latent heterogeneity in activity intensity. Such a framework could be extended to support urban planning decisions, for example by identifying regions whose statistical structure indicates vulnerability or resilience to exogenous shocks.\\
Importantly, this work contributes to bridging the gap between urban form, the spatial disposition of POIs, and its function (the POIs themselves). Understanding this interplay is crucial for characterizing how cities function: POIs do not exist in isolation, but rather compete for attention, interact through flows of people and resources, and coexist in patterns that reflect both complementarities and constraints. By modeling the emergence of spatial structure from locally homogeneous intensities that combine into global heterogeneity, our approach offers a step toward the characterization of these dynamics and overall towards an empirically grounded theory of urban organization.\\
The empirical validation on detailed data from the city of Bologna confirms that these theoretical results are not merely an abstract construct but can be operationalized to interpret real-world urban patterns with remarkable accuracy. Building on this success, future developments could explore whether the same mechanism governs spatial scaling in other cities and contexts. One promising avenue is the temporal extension of this model to investigate how the inferred latent probabilities and the spatial configuration of Poissonian regions evolve over time, potentially uncovering dynamics associated with urban growth, economic transitions, or infrastructure changes. Another direction concerns the integration of additional covariates, such as socio-economic indicators or mobility data, to enrich the interpretation of the observed scaling patterns and to understand how different urban functions co-evolve.\\
Overall, by bridging analytical derivation and empirical validation, this work contributes a tractable and generalizable explanation for the widespread presence of scaling laws in urban systems, opening new pathways for both theoretical exploration and applied research in spatial complexity.

\section*{Author Contributions}
E.A. conceived and designed the study, carried out the mathematical formulation and calculations, performed the data analysis, and wrote the manuscript. M.N. contributed through scientific discussions. U.M. and R.G. contributed through scientific discussions and manuscript review. All authors read and approved the final version of the paper.

\section*{Acknowledgments}

 E.A. and M.N. acknowledges the financial support received from the TALEA Green Cells Leading the Green Transition (EUI02-064) project. R.G. and M.N. acknowledge the support of the PNRR ICSC National Research Centre for High Performance Computing, Big Data and Quantum Computing (CN00000013), under the NRRP MUR program funded by the NextGenerationEU.

\bibliography{biblio}
\appendix
\section*{Appendix: Additional Figures}

\begin{figure*}[h!]
\centering
\includegraphics[width=.6\textwidth]{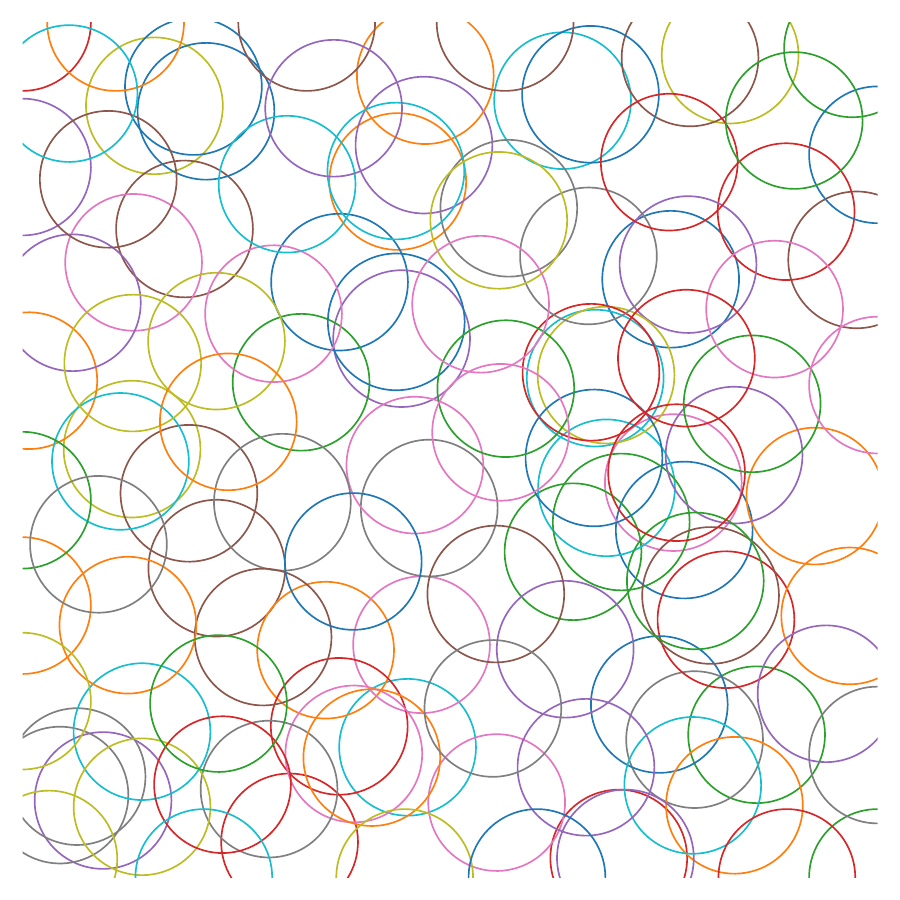}
\caption{Toy-model illustration of the ball covering procedure. The urban surface is first covered by a random Boolean model of fixed-radius balls, and the remaining uncovered gaps are subsequently filled through a greedy covering step, ensuring complete coverage of the study area.}
\label{fig:covering}
\end{figure*}

\begin{figure*}[h!]
\centering
\subfigure[Business and Professional Services]{\includegraphics[width=.55\textwidth]{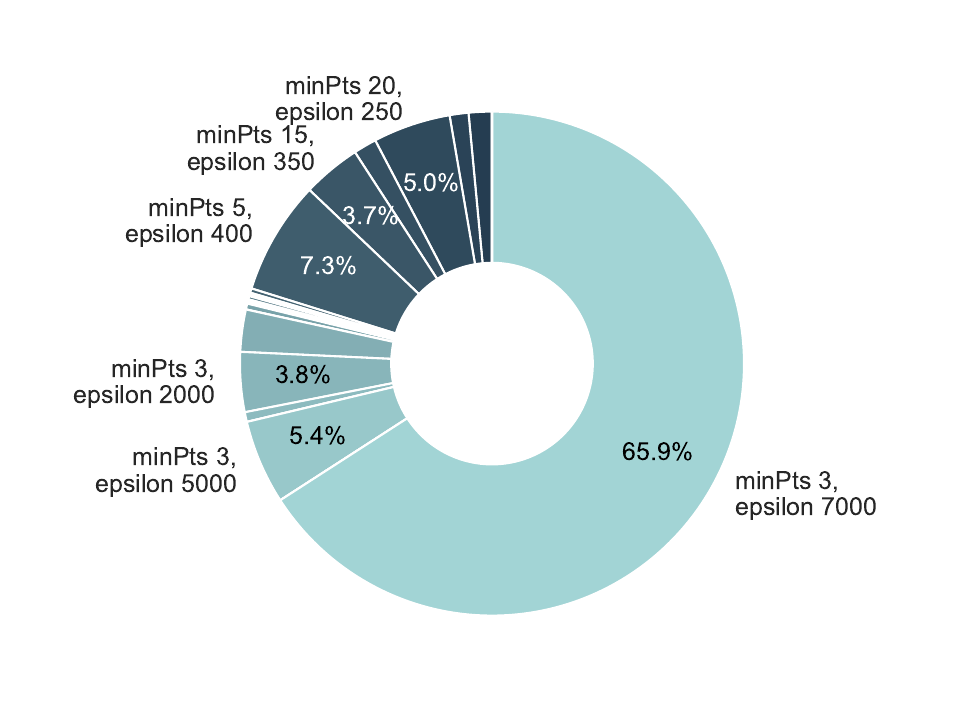}}
\subfigure[Community and Government]{\includegraphics[width=.55\textwidth]{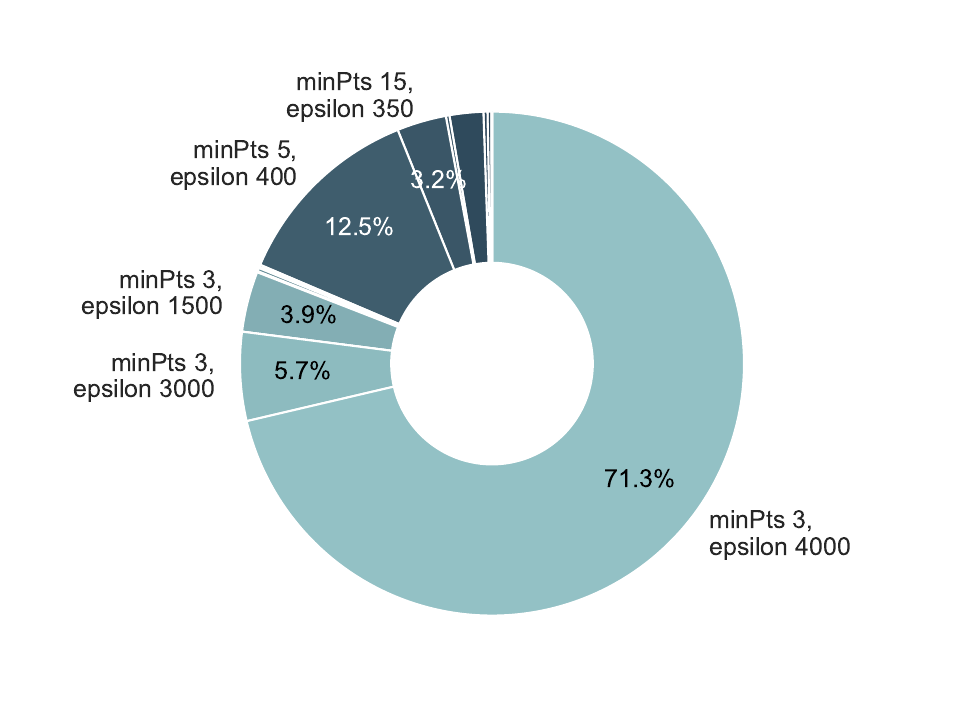}}
\subfigure[Dining and Drinking]{\includegraphics[width=.55\textwidth]{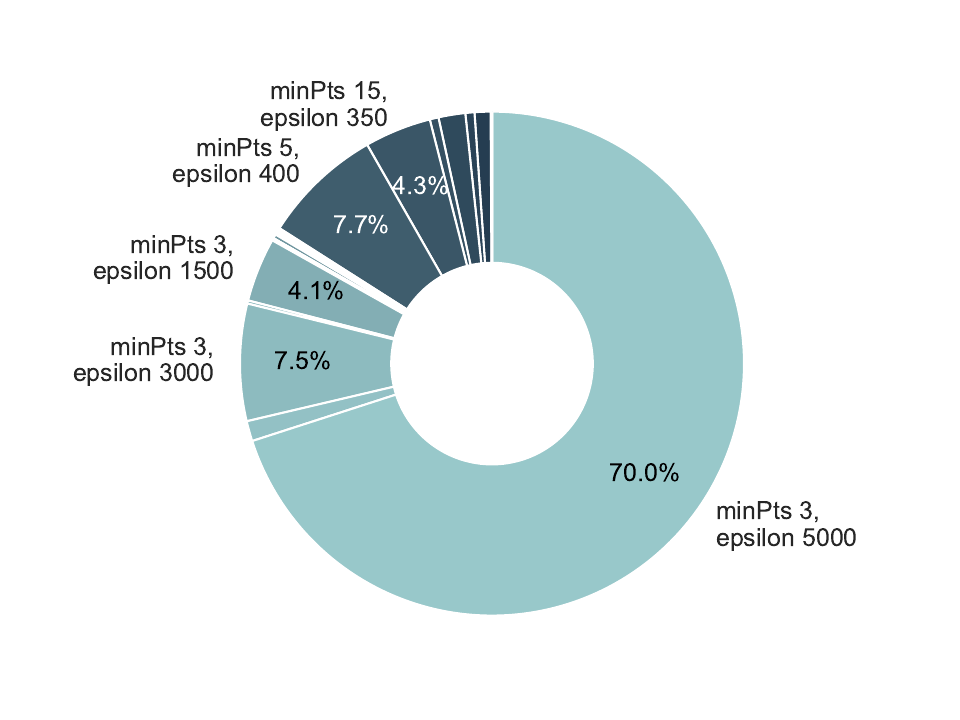}}
\caption{Pie charts showing the proportion of Bologna’s total urban surface occupied by POI clusters, identified through density-based spatial clustering (DBSCAN), per category.}\label{fig:torte1}
\end{figure*}

\begin{figure*}[h!]
\centering
\subfigure[Health and Medicine]{\includegraphics[width=.55\textwidth]{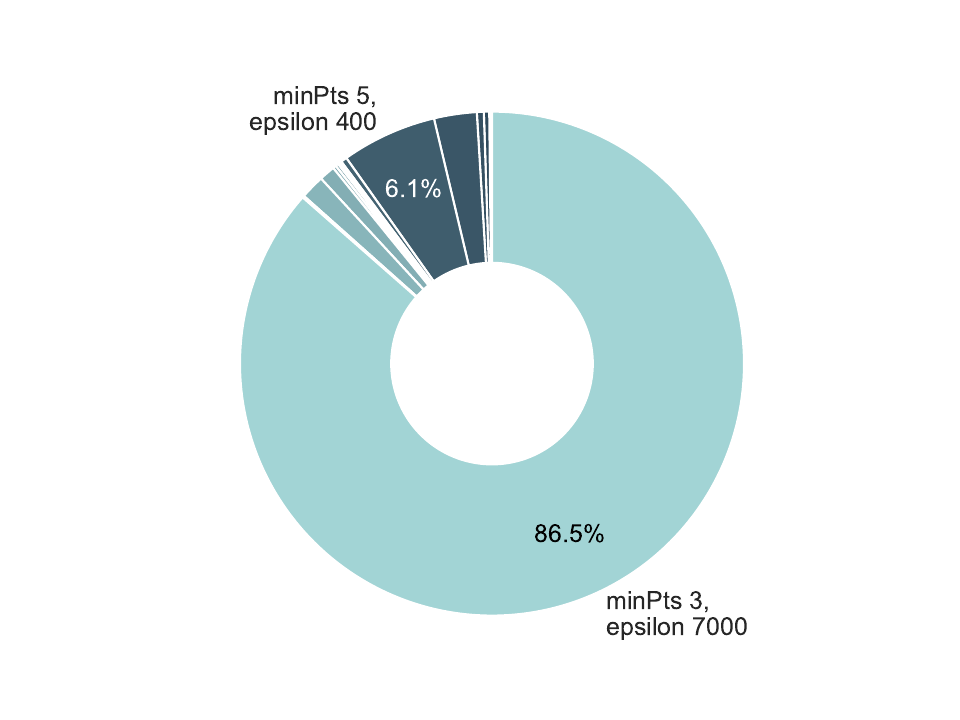}}
\subfigure[Retail]{\includegraphics[width=.55\textwidth]{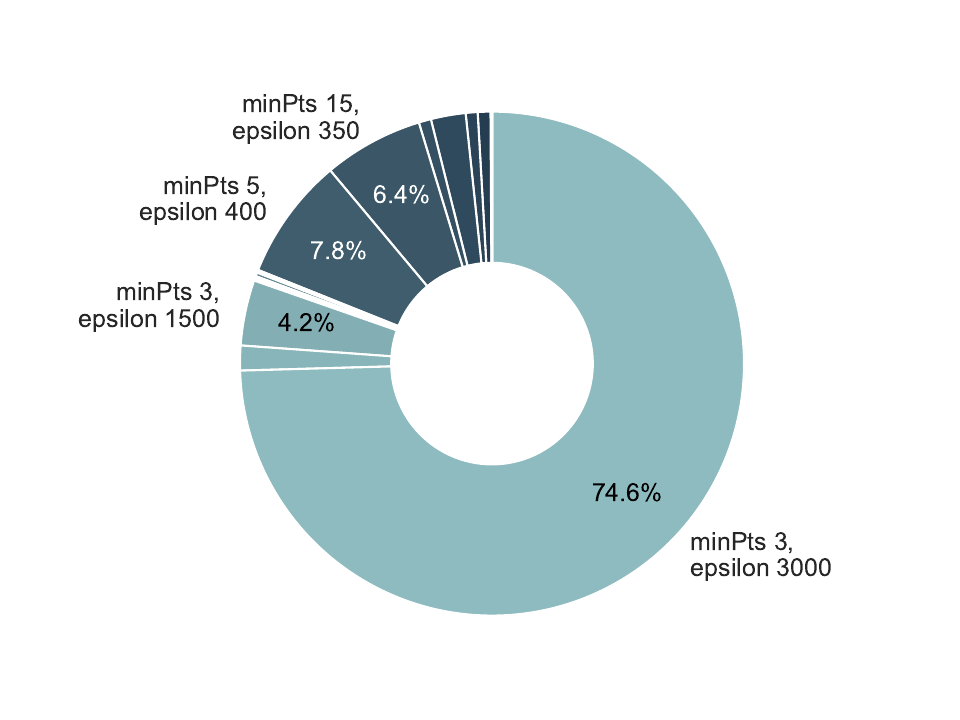}}
\subfigure[Travel and Transportation]{\includegraphics[width=.55\textwidth]{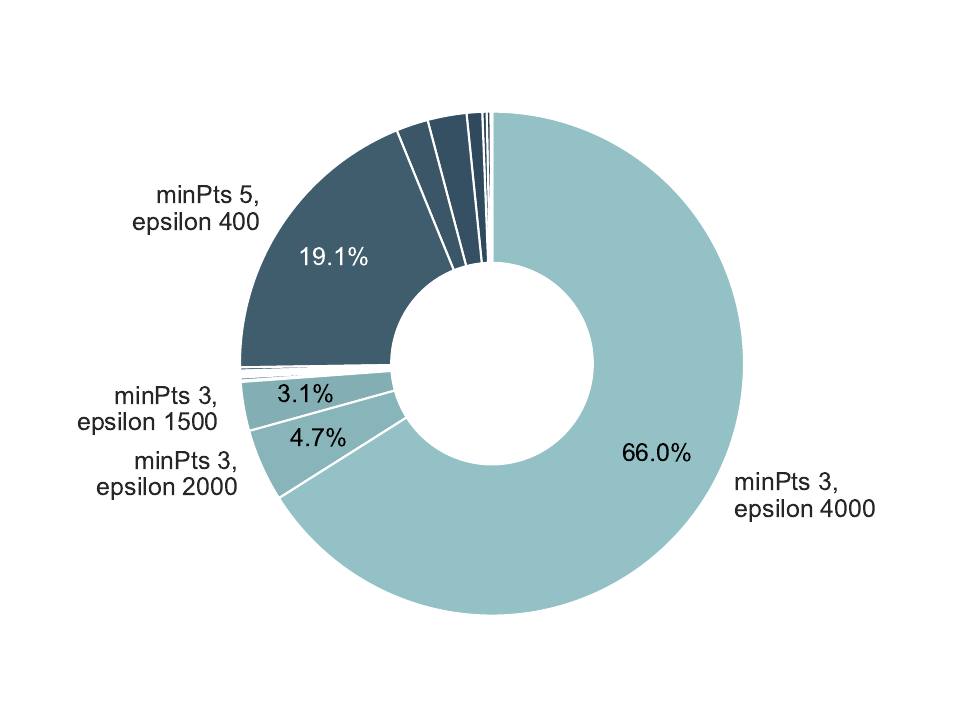}}
\caption{Pie charts showing the proportion of Bologna’s total urban surface occupied by POI clusters, identified through density-based spatial clustering (DBSCAN), per category.}\label{fig:torte2}
\end{figure*}

\begin{figure*}[h!]
\centering
\subfigure[Business and Professional Services]{\includegraphics[width=.55\textwidth]{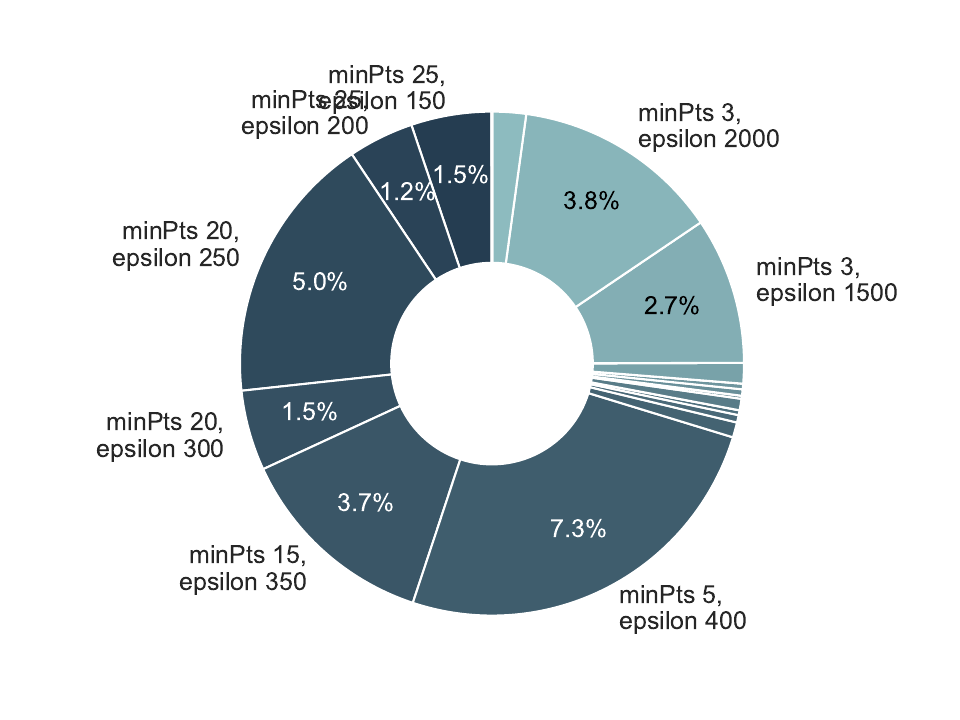}}
\subfigure[Community and Government]{\includegraphics[width=.55\textwidth]{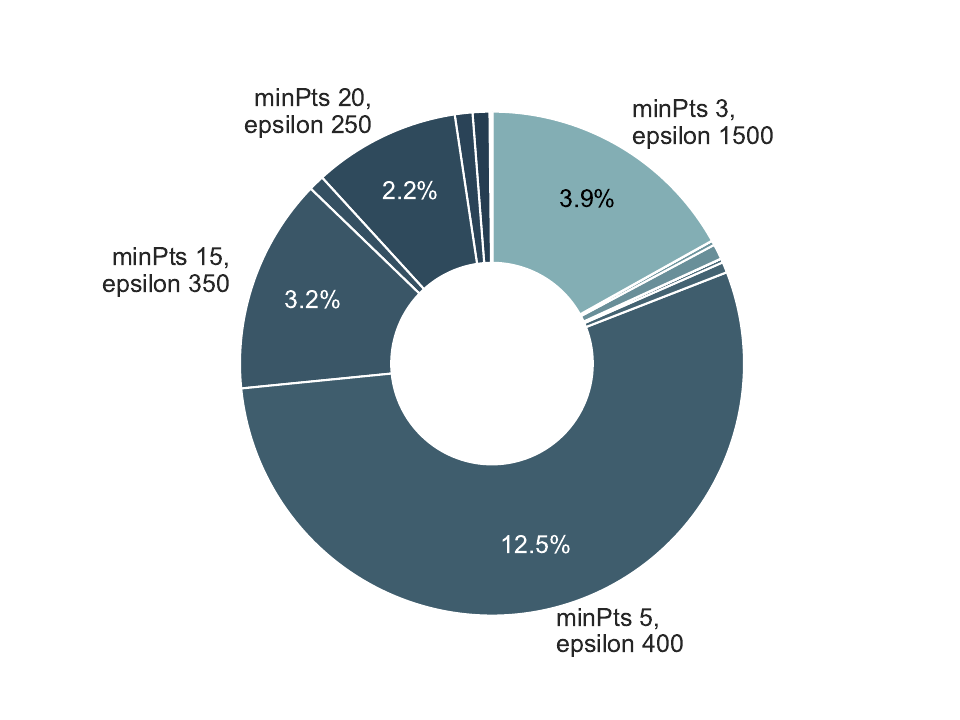}}
\subfigure[Dining and Drinking]{\includegraphics[width=.55\textwidth]{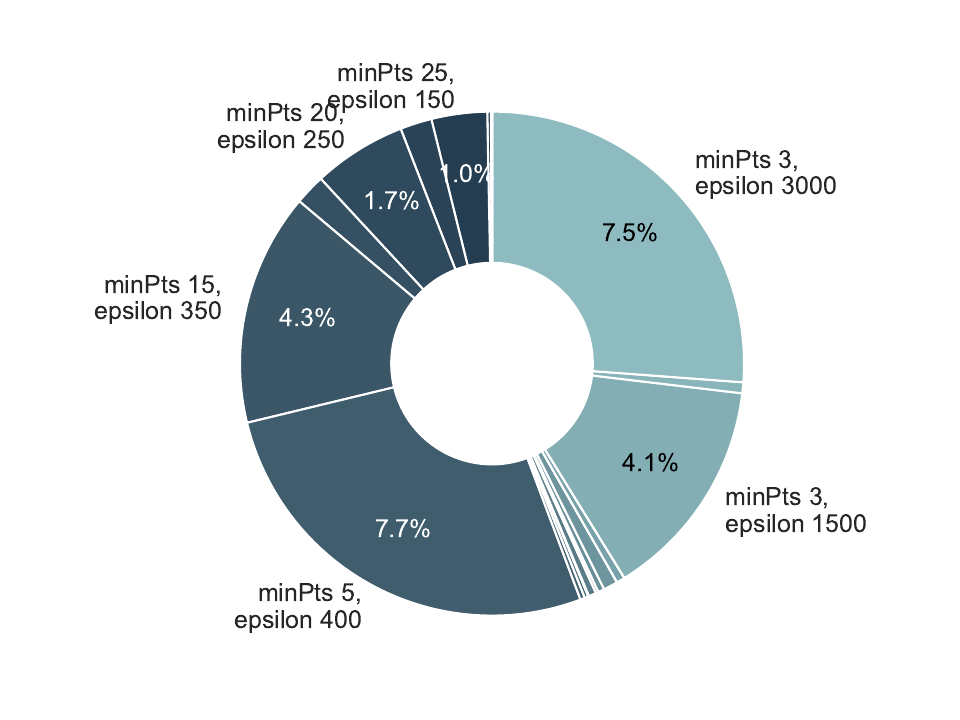}}
\caption{Zoomed view of non-dominant POI clusters’ surface area per category.}\label{fig:tortezoom1}
\end{figure*}

\begin{figure*}[h!]
\centering
\subfigure[Health and Medicine]{\includegraphics[width=.55\textwidth]{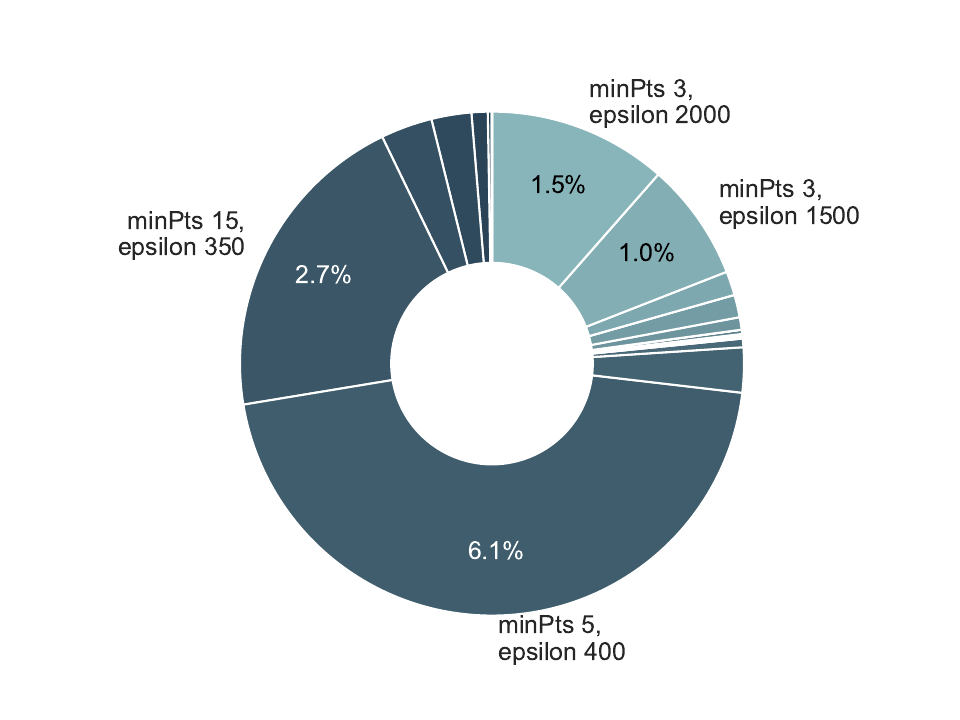}}
\subfigure[Retail]{\includegraphics[width=.55\textwidth]{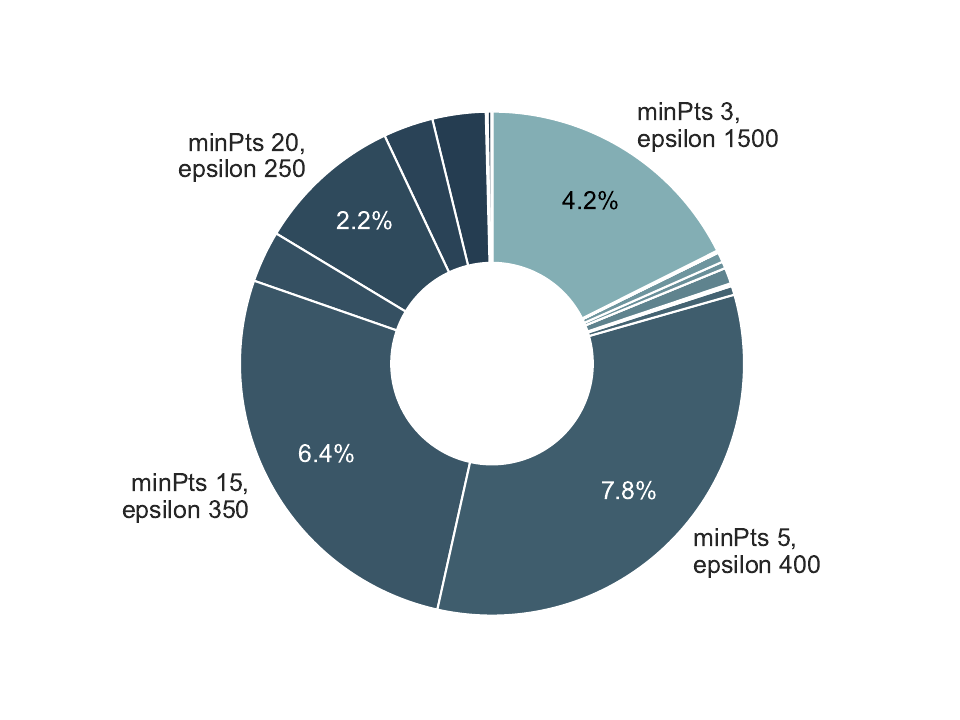}}
\subfigure[Travel and Transportation]{\includegraphics[width=.55\textwidth]{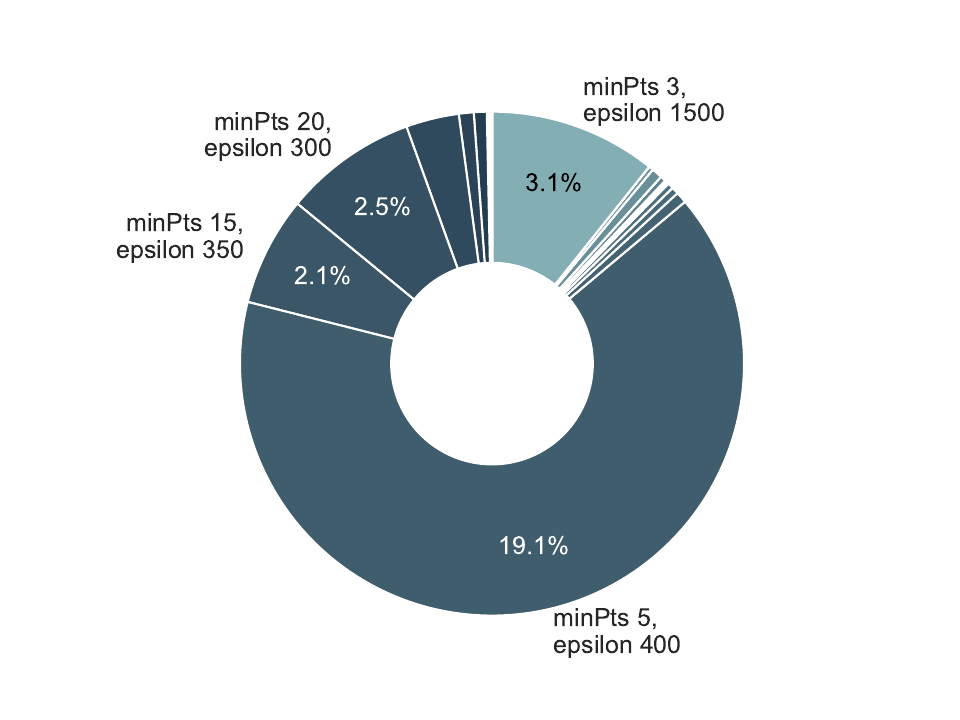}}
\caption{Zoomed view of non-dominant POI clusters’ surface area per category.}\label{fig:tortezoom2}
\end{figure*}

\end{document}